\documentclass[aps,prd,preprint,superscriptaddress,showpacs,floatfix,nobibnotes,nofootinbib]{revtex4-1}
\usepackage[usenames, dvipsnames]{color}

\usepackage{float}

\usepackage{latexsym}
\usepackage{amsmath}
\usepackage{amssymb}
\usepackage{graphicx}
\usepackage{setspace}
\usepackage{longtable}
\usepackage{enumitem}
\usepackage[bottom]{footmisc}
\usepackage{slashed}

\usepackage{bm}
\usepackage{epsfig}
\newcommand{\bea}{\begin{eqnarray}}
\newcommand{\eea}{\end{eqnarray}}

\newcommand{\nn}{\nonumber}

\begin{document}
\setlength{\baselineskip}{18pt}

\title{Collective modes in anisotropic systems}

\author{Margaret E. Carrington}
\affiliation{Department of Physics and Astronomy, Brandon University,
Brandon, Manitoba R7A 6A9, Canada}
\affiliation{Winnipeg Institute for Theoretical Physics, Winnipeg, Manitoba, Canada}

\author{Bailey M. Forster}
\affiliation{Department of  Physics and Astronomy, Brandon University,
Brandon, Manitoba R7A 6A9, Canada}

\author{Sofiya Makar}
\affiliation{Department of Mathematics, Brandon University,
Brandon, Manitoba R7A 6A9, Canada}

\date{October 05, 2021}

\begin{abstract}
We study  collective modes in anisotropic plasmas of quarks and gluons using a quasi-particle picture and a hard loop approximation. 
We use a general class of anisotropic distribution functions, and we consider chirally asymmetric systems.
We introduce a complete tensor basis to decompose the gluon polarization tensor into a set of nine scalar functions. 
We derive and solve the corresponding dispersion equations. 
Imaginary modes are particularly important because of their potential influence on plasma dynamics. 
We explore in detail their dependence on the chiral chemical potential and the parameters that characterise the anisotropy of the system. 
We show that our generalized distributions produce dispersion relations that are much richer in structure than those obtained with a simple one parameter deformation of an isotropic distribution. 
In addition, the size and domain of the imaginary solutions are enhanced, relative to those obtained with a one parameter deformation. 
Finally, we show that the influence of even a very small chiral chemical potential is significantly magnified when anisotropy is present. 

\end{abstract}

\maketitle

\section{Introduction}

We consider a relativistic plasma of charged particles that is not necessarily in equilibrium, but is in the regime where a quasi-particle description is valid.
We use an effective phase-space description in terms of distribution functions that are not assumed thermal or isotropic. 
A convenient parametrization for the distribution function of a spheroidially anisotropic system was introduced in \cite{Romatschke:2003ms,Romatschke:2004jh}. 
The idea is to introduce one parameter that effectively stretches or squeezes the spherically symmetric isotropic distribution in one direction. 
This parametrization is interesting in the context of heavy ion collisions where one expects the parton distribution that is produced by the collision to be squeezed along the beam axis. 
Dispersion relations are obtained from the poles of the dressed propagators of the partonic quasi-particles. 
One finds that the spectrum of collective excitations is much richer than the spectrum of a system in thermal equilibrium. 
It is especially interesting that there are imaginary modes, which are associated with instabilities and could play an important role in the thermalization of the plasma \cite{Mrowczynski:1993qm,Arnold:2003rq, Kurkela:2011ti,Kurkela:2011ub,Ipp:2010uy,Attems:2012js}. 
Anisotropic plasmas have been studied intensively in the context of quark-gluon plasmas, which thermalize earlier than expected, for reasons that are as yet unclear. 
Collective modes have been calculated for both gluons \cite{Romatschke:2003ms,Romatschke:2004jh,Carrington:2014bla} and quarks \cite{Schenke:2006yp}. 
An ellipsoidal generalization was developed in \cite{Kasmaei:2016apv,Kasmaei:2018yrr}. 
These distributions have been used to study various properties of quark-gluon plasma (QGP) including 
transport coefficients \cite{Baier:2008js,Romatschke:2004au,Carrington:2015xca,Carrington:2016mhd}, 
heavy quark bound states \cite{Dumitru:2007hy,Burnier:2009yu,Dumitru:2009fy,Boguslavski:2021baf}, 
and photon and dilepton production \cite{Schenke:2006yp,Martinez:2008di,Hauksson:2020wsm,Kasmaei:2018oag,Kasmaei:2019ofu}.

Isotropic plasmas also support imaginary collective modes, if there is a chiral imbalance \cite{Akamatsu:2013pjd}. 
Chiral plasmas with spheroidal anisotropy were studied in Ref. \cite{Kumar:2014fta}. 
Chiral systems are of interest in a wide variety of different situations in particle physics, nuclear physics, condensed matter physics, and cosmology. 
One important example is the chiral magnetic effect (CME) which is the production of a parity violating current in a plasma, in the presence of a magnetic field and an asymmetry between left and right handed fermions \cite{Fukushima:2008xe}. 
A chirally asymmetric plasma can be described  using a chiral chemical potential defined as $\mu_5 \equiv (\mu_R-\mu_L)/2$, where $\mu_L$ and $\mu_R$
are the chemical potentials of the left and right handed fermions. 
The CME is particularly interesting in the context of heavy-ion collisions because it has been argued that the quark gluon plasma that is produced in non-central collisions may contain regions where $\mu_5$ is locally finite \cite{Kharzeev:1998kz, Kharzeev:2001ev,Kharzeev:2004ey,Kharzeev:2007jp,Kharzeev:2009fn}. 
Instabilities in chiral plasmas have also been studied in electroweak theory at large lepton chemical potential \cite{redlich-1985} and in the context of the early universe at $T\gg\mu_5$ \cite{Laine:2005bt}. 

Collective modes can be studied using either kinetic theory, or effective field theories. 
The equivalence of the kinetic theory method and a hard thermal loop (HTL) effective theory approach, for isotropic and chirally symmetric systems, was shown in \cite{Blaizot:1993zk,Kelly:1994ig,Litim:2001db}. 
For anisotropic chirally symmetric systems, the connection between kinetic theory and a generalization of the HTL effective theory which is called a hard loop (HL) effective theory has also been established \cite{Mrowczynski:2000ed,Mrowczynski:2004kv}. 
Isotropic chiral kinetic theories have been developed in \cite{Son:2012wh,Stephanov:2012ki,Son:2012zy,Manuel:2013zaa,Carignano:2018gqt}, and applied to a variety of physical problems, see for example \cite{Manuel:2015zpa,Carignano:2018thu,Carignano:2021mrn}.

In this paper we use anisotropic distributions that are more general than those used in  previous works, and we study the spectrum of collective modes in anisotropic plasmas with non-zero chiral chemical potential. 
The method can be applied generally to either a QED plasma of ultrarelativistic electrons and
positrons or a QGP. 
%
We use natural units where $\hbar = c =1$. The indices $i,j,k = 1, 2, 3$ and $\mu, \nu = 0, 1, 2, 3$ label, respectively, the Cartesian spatial coordinates and those of Minkowski space. Our metric is mostly minus $g_{\mu\nu} = (1,-1,-1,-1)_{\rm diag}$. 
We use capital letters for four-vectors so that, for example, $P^2=p_0^2-\vec p\cdot\vec p =  p_0^2-p^2$. We define the unit vector $\hat p = \vec p/|\vec p|$ and we will also use $\hat p_0 = p_0/p$. Real solutions to dispersion equations will be denoted $\omega(\vec p)$ and imaginary solutions are written $i\gamma(\vec p)$.

\section{Isotropic formalism}
\label{iso-sec}

In vacuum the photon polarization tensor can be written in terms of one scalar function using a transverse projection operator as $\Pi^{\mu\nu}(p_0,\vec p) = (P^2 g^{\mu\nu}-P^\mu P^\nu)\,\Pi(p_0,\vec p)$. 
At finite temperature the rest frame of the heat bath, which we define with the four-vector $n^\mu = (1,0,0,0)$, breaks Lorentz invariance.
Time-like axial gauge (TAG) is  particularly useful at finite temperature, because the gauge condition is imposed in the heat bath rest frame. 
In TAG we only have to consider the spatial components of the propagator, for which the Dyson equation has the form
\bea
D^{ij -1}(p_0,\vec p) = D_0^{ij -1}(p_0,\vec p) -\Pi^{ij}(p_0,\vec p) = P^2\delta^{ij} + p^2 \hat p^i \hat p^j  -\Pi^{ij}(p_0,\vec p)\,.
\label{dyson-def}
\eea

In an equilibrated chirally symmetric plasma, the polarization tensor is symmetric and has two independent components, one transverse and the other longitudinal with respect to the three-vector $\vec p$. 
An isotropic system with finite chiral chemical potential was first considered in \cite{Akamatsu:2013pjd}, and we review the formalism presented in that paper below. 
In the presence of finite $\mu_5$ the polarization tensor has an additional asymmetric component and can be written
\bea
\Pi^{ij}(p_0,\vec p) = (\delta^{ij}-\hat p^i\hat p^j) \Pi_T(p_0,\vec p) + \hat p^i\hat p^j \Pi_L(p_0,\vec p) + i\epsilon^{ijm}\hat p^m \Pi_A(p_0,\vec p) \,. \label{pi-aka}
\eea
The propagator, obtained from inverting the Dyson equation, is
\bea
D^{ij}(p_0,\vec p) &=& (\delta^{ij}-\hat p^i\hat p^j) \frac{(P^2-\Pi_T(p_0,\vec p))}{(P^2-\Pi_T(p_0,\vec p))^2 - \Pi_A^2(p_0,\vec p)} + \hat p^i\hat p^j\frac{1}{p_0^2-\Pi_L(p_0,\vec p)}  \nn \\[2mm]
&+& i\epsilon^{ijm}\hat p^m \frac{\Pi_A(p_0,\vec p)}{(P^2-\Pi_T(p_0,\vec p))^2 - \Pi_A^2(p_0,\vec p)}\,.
\label{prop-aka}
\eea
The dispersion relations are obtained from the dispersion equations which give the poles of the retarded propagator. From equation (\ref{prop-aka}) we have that the dispersion equations for an isotropic chiral plasma are
\bea
&& p_0^2-\Pi_L(p_0,\vec p) = 0 \label{Lmodes} \\ 
&& P^2-\big(\Pi_T(p_0,\vec p) + \Pi_A(p_0,\vec p) \big) = 0 \label{Pmodes} \\
&& P^2- \big(\Pi_T(p_0,\vec p) - \Pi_A(p_0,\vec p) \big) = 0 \,.\label{Mmodes} 
\eea
Throughout this paper we refer always to the retarded polarization tensor, and we will not include any subscripts to indicate this.

Our notation for the equilibrium distribution function is (see equation (\ref{distro-def}))
\bea
n(k) =n^+(k) =  \frac{1}{e^{\beta(k-\mu)}+1} \text{~~and~~} \bar n(k) = n^-(-k) =  \frac{1}{e^{\beta(k+ \mu)}+1}\, . \label{nn-def}
\eea
The distributions for right/left handed particle/anti-particles are written $n_R(k)$, $\bar n_R(k)$, $n_L(k)$ and $\bar n_L(k)$, where, for example,  $n_R(k) = n(k)\big|_{\mu=\mu_R}$. 

We calculate the 1-loop photon polarization tensor in the HTL approximation (see Appendix \ref{retarded-all} for details). 
The contribution from right handed fermions is
\bea
&& \Pi^{ij}_R(p_0,\vec p) = \Pi^{ij}_{{R\rm even}}(p_0,\vec p) + \Pi^{ij}_{R{\rm odd}}(p_0,\vec p) \nn\\[2mm]
&& \Pi^{ij}_{R{\rm even}}(p_0,\vec p) = 2g^2\int\frac{d^3 k}{(2\pi)^3}\frac{n_R(k)+\bar n_R(k)}{k} 
\left(\delta^{ij}+\frac{v^i p^j + p^i v^j}{P\cdot V +i\epsilon}- \frac{P^2 v^i v^j}{(P\cdot V+i\epsilon)^2}\right) \label{pi-even} \\
&& \Pi^{ij}_{R{\rm odd}}(p_0,\vec p) =  i g^2 P^2 \epsilon^{ijm} \int\frac{d^3 k}{(2\pi)^3}\frac{n_R(k) - \bar n_R(k)}{k^2} 
\frac{(p_0 v^m - p^m)}{(P\cdot V +i\epsilon)^2} \, \label{pi-odd}
\eea
and the corresponding expression for left handed fermions is obtained from the transformation $\mu_R\to \mu_L$. 
We use $P\cdot V = p_0-\vec p\cdot\vec v$ and  
$\vec k/\sqrt{k^2+m^2} \approx \hat k \equiv \vec v$ since massless fermions are consistent with the hard loop approximation. 
It is straightforward to modify the HTL integrals to the case of a QCD plasma by changing the QED coupling constant to the QCD one, and using \cite{Le-Bellac-2000} 
\bea
&& n(k)+\bar n(k) \to \frac{N_f}{2}(n_q(k)+\bar n_q(k)) + N_c n_g(k) \nn\\
&& n(k)-\bar n(k) \to \frac{N_f}{2}(n_q(k)- \bar n_q(k)) 
\eea
where the subscripts $q$ and $g$ refer to quark (or anti-quark) and gluon distribution functions. 
Equations (\ref{pi-even}, \ref{pi-odd}) can be rewritten in a form that is sometimes more useful by integrating by parts. 
A straightforward calculation gives
\bea
\Pi^{ij}_{R{\rm even}}(p_0,\vec p) &=&  - 2 g^2 \int \frac{d^3 k}{(2\pi)^3} \frac{\partial (n_R(k)+\bar n_R(k))}{\partial k^m}  v^i \left(\delta^{jm} + \frac{v^j p^m}{P\cdot V} \right)
\label{pi-even-2}\\
\Pi^{ij}_{R{\rm odd}}(p_0,\vec p) &=&  - i g^2 \epsilon^{ijm} \int \frac{d^3k}{(2\pi)^3} \frac{1}{k} \frac{\partial}{\partial k^l}(n_R(k) - \bar n_R(k))
 \left(p_0 \delta^{lm} + \frac{(p_0 v^m-p^m)p^l}{P\cdot V}
\right)\,. ~~~~~~~~~~~ \label{pi-odd-2}
\eea

From the definitions in equations (\ref{pi-rl}, \ref{pi-sum}) one finds 
\bea
\Pi^{ij}(p_0,\vec p) &=& \frac{1}{2}\left(\Pi^{ij}_R(p_0,\vec p) + \Pi^{ij}_L(p_0,\vec p)\right) \nn \\
 &=& \frac{1}{2}\left(\Pi^{ij}_{R{\rm even}} + \Pi^{ij}_{L{\rm even}}\right) + \frac{1}{2}\left(\Pi^{ij}_{R{\rm odd}} + \Pi^{ij}_{L{\rm odd}}\right) \nn \\
 &=& \frac{1}{2}\left(\Pi^{ij}_{R{\rm even}} + \Pi^{ij}_{R{\rm even}}\big|_{\mu_R \to \mu_L}\right) + \frac{1}{2}\left(\Pi^{ij}_{R{\rm odd}} + \Pi^{ij}_{R{\rm odd}}\big|_{\mu_R \to -\mu_L}\right) \,.\label{combo}
\eea
The functions $\Pi_T(p_0,\vec p)$, $\Pi_L(p_0,\vec p)$ and $\Pi_A(p_0,\vec p)$ in equations (\ref{pi-aka}, \ref{prop-aka}) are calculated by applying the projection operators 
${\cal P}_T^{ij} = (\delta^{ij}-\hat p^i\hat p^j)/2$, ${\cal P}_L^{ij} = \hat p^i\hat p^j$, and ${\cal P}_A^{ij} = -i\epsilon^{ijm}\hat p^m/2$ to 
$\Pi^{ij}(p_0,\vec p)$. 
The resulting expressions for the three scalar components of the polarization tensor are
\bea
&& \Pi_T(p_0,\vec p) = m_D^2 \frac{p_0^2}{2p^2} \left(1 - \frac{P^2}{2p_0 p} \ln \left(\frac{p_0+p+i\epsilon}{p_0- p+i\epsilon}\right) \right) \nn\\
&& \Pi_L(p_0,\vec p) = - m_D^2 \frac{p_0^2}{p^2}\left( 1- \frac{p_0}{2p} \ln\left(\frac{p_0+p+i\epsilon}{p_0- p+i\epsilon}\right) \right) \nn\\
&& \Pi_A(p_0,\vec p) = -g^2 \frac{\mu_5 P^2 }{2\pi^2 p }\left(1-\frac{p_0}{2p} \ln\left(\frac{p_0+p+i\epsilon}{p_0- p+i\epsilon}\right) \right) \label{res-iso}
\eea
where we have used $\mu_5 = (\mu_R-\mu_L)/2$ and defined the Debye mass parameter
\bea
m_D^2 &=& 2g^2 \int \frac{d^3k}{(2\pi)^3} \frac{1}{k} \left(\left[n(k)+\bar n(k)\right]_{\mu_R} + \left[n(k)+\bar n(k)\right]_{\mu_L} \right)  = g^2 \left(\frac{T^2}{3} + \frac{1}{2\pi^2}(\mu_R^2+\mu_L^2)\right)\,.\nn\\ \label{def-mD}
\eea
The transverse and longitudinal components of the polarization tensor are the familiar HTL results, and in the chirally symmetric limit $\mu_R = \mu_L$ gives $\mu_5=0$ and therefore $\Pi_A(p_0,\vec p)=0$. 
Throughout the rest of this paper we use the shorthand notation $\hat\mu_5 = g^2\mu_5/(2\pi^2)$. 
In all of our numerical calculations we set the Debye mass to one, which is equivalent to defining all dimensionful quantities in units of the Debye mass. 

\begin{figure}[htb]
\begin{center}
\includegraphics[width=12.5cm]{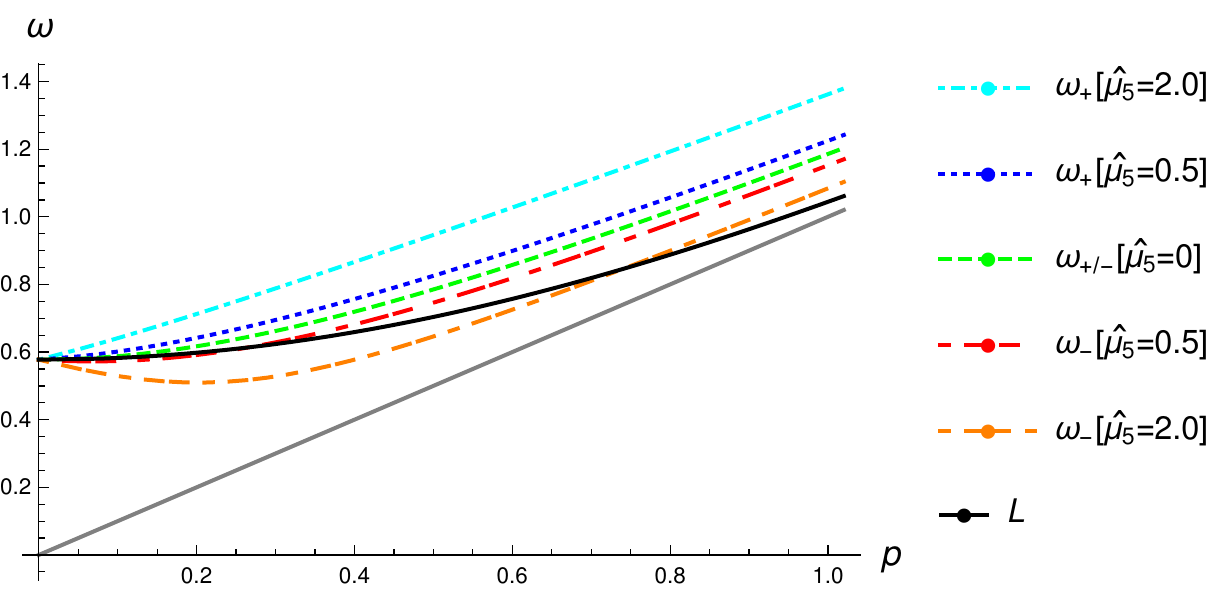}
\includegraphics[width=12.5cm]{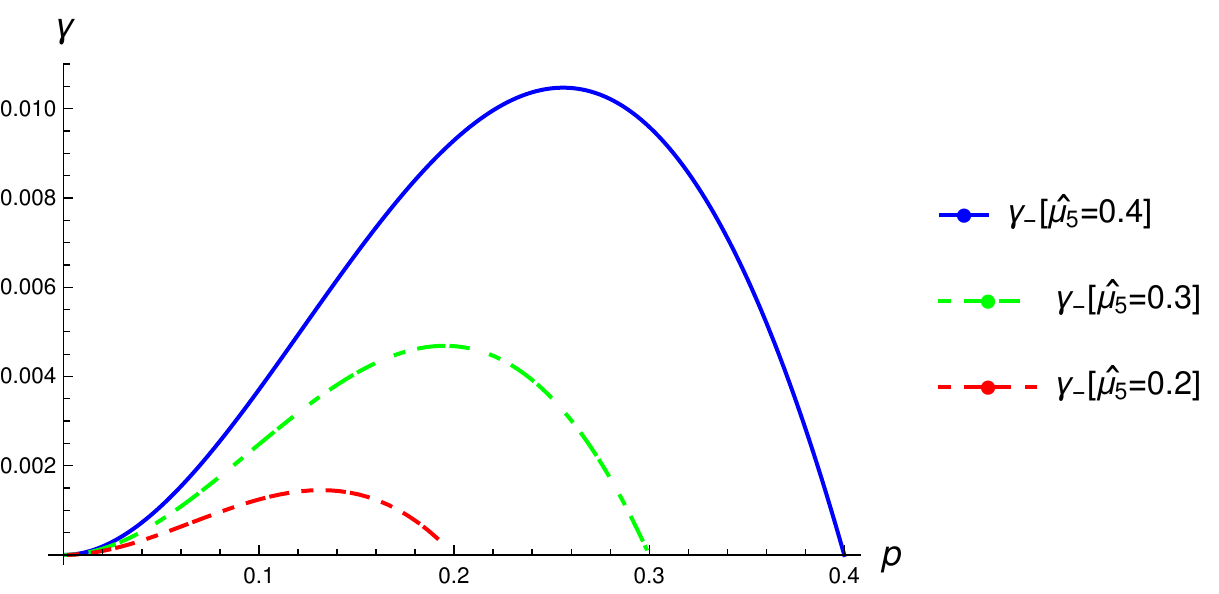}
\end{center}
\caption{Real ($\omega$) and imaginary ($\gamma$) solutions to the dispersion equations (\ref{Lmodes}-\ref{Mmodes}) for different values of $\hat\mu_5 = g^2\mu_5/(2\pi^2)$. The units are defined by setting $m_D=1$. The diagonal gray line in the top figure is the light-cone.\label{disp-iso-fig}}
\end{figure}

The dispersion relations are obtained by solving equations (\ref{Lmodes}-\ref{Mmodes}) 
where the functions $\Pi_L(p_0,\vec p)$, $\Pi_T(p_0,\vec p)$ and $\Pi_A(p_0,\vec p)$ are given in equation (\ref{res-iso}). 
There are pure real and pure imaginary solutions that appear in positive and negative pairs. 
Equation (\ref{Lmodes}) has two real solutions which are just the usual longitudinal HTL modes, and are written $\pm\omega_L(p)$. 
Equation (\ref{Pmodes}) also has two real solutions, which we denote $\pm\omega_+(p)$. 
Equation (\ref{Mmodes}) has two real solutions, which we call $\pm\omega_-(p)$, and two extra solutions that appear at values of the wave vector below the critical value $p_{\rm crit}=\hat\mu_5$. 
These solutions are pure imaginary and denoted $\pm i\gamma_-(p)$. 
The critical wave vector at which the imaginary solutions appear can be found with a Nyquist analysis (see  Appendix \ref{nyquist-sec} for details). 
When $\hat\mu_5=0$ we have $\omega_+(p) = \omega_-(p) = \omega_T(p)$ where $\omega_T(p)$ is the transverse HTL mode. 
In figure \ref{disp-iso-fig} we show solutions to equations (\ref{Lmodes} - \ref{Mmodes}) for different values of the parameter $\hat\mu_5$. 

\section{Anisotropic formalism}
\subsection{Distribution function}
\label{Hdef-section}

Next we consider the situation that the momentum distribution is not isotropic. 
A simple way to account for momentum anisotropy is to replace the distribution function in equations (\ref{pi-even}, \ref{pi-odd}) using $n(k) \to n(\vec k) = C_\xi \, n\big(k H_\xi(\vec v)\big)$ where $\vec v = \vec k/k$, and similarly for $\bar n(k)$. The subscript $\xi$ indicates dependence on a set of anisotropy parameters that can be used to construct a distribution that is deformed relative to the isotropic one. The factor $C_\xi$ is a normalization and can be defined in different ways depending on the calculation being done. Our choice is explained at the end of this section.  

We can construct a completely general expression for the function $H_\xi(\vec v)$ as a sum of terms that are products of anisotropy parameters and dot products of the vector $\vec v$ with two perpendicular unit vectors, which we will write $\hat n_3$ and $\hat n_1$. 
We are primarily interested in distributions of partons produced in heavy-ion collisions. In this context, we will take $\hat n_3$ to define the beam axis and $\hat n_1$ to give the direction of transverse anisotropy. 
We restrict to functions that satisfy the condition $H_\xi(\vec v) = H_\xi(-\vec v)$ and use an expression of the form
 \bea
 H^2_\xi(\vec v)  &=& \xi_0
 +\xi _2(\vec n_1\cdot\vec v)^2+\xi _9 (\vec n_3\cdot\vec v)^2
 +\xi _6 (\vec n_1\cdot\vec v)(\vec n_3\cdot\vec v) \label{H-def} \\
     && \hspace*{-1.2cm} 
     +\xi _4 (\vec n_1\cdot\vec v)^4
      + \xi _8 (\vec n_1\cdot\vec v)^3 (\vec n_3\cdot\vec v)
 +\xi _{11}(\vec n_1\cdot\vec v)^2 (\vec n_3\cdot\vec v)^2
     +\xi _{13} (\vec n_1\cdot\vec v)(\vec n_3\cdot\vec v)^3  
     +\xi _{14} (\vec n_3\cdot\vec v)^4 \,. \nonumber
 \eea
 When $\xi_0=1$ and $\xi_{i\ne 0}=0$, we have $H^2_\xi(\vec v)=1$,  and the distribution is isotropic. 
For an arbitrary choice of the anisotropy parameters $\xi_i$, the isotropic distribution is expanded in the direction of $\vec v$ if $H_\xi(\vec v)<1$, and contracted if $H_\xi(\vec v)>1$. 
The values of the anisotropy parameters must be chosen so that $H^2_\xi(\vec v)$ is positive for all orientations of the vector $\vec v$, 
which is equivalent to the requirement that $H_\xi(\vec v)$, and therefore the argument of the distribution function, is real and positive.
In section \ref{coord} we will show that our expression (\ref{H-def}) is equivalent to including spherical harmonics up to second order. 

Equation (\ref{H-def}) includes the parametrizations used in several previous calculations.
In the original work of Ref. \cite{Romatschke:2003ms} the authors used $\xi_0=1$ and $\xi_9\in(-1,\infty)$ with  all other parameters set to zero. 
The chosen value of $\xi_9$ allows one to consider either a slightly prolate momentum distribution ($-1<\xi_9<0$), which is elongated along the beam axis, or an arbitrarily oblate distribution ($\xi_9>0$), which  is squeezed in the direction of the beam axis. 
In the calculations of Ref. \cite{Kasmaei:2016apv,Kasmaei:2018yrr}, where ellipsoidal asymmetry was included for the first time, 
the authors used $\xi_0=1$ and various values of $\xi_2$ and $\xi_9$ with all other parameters zero. 
In Ref. \cite{Carrington:2014bla} the authors considered a one parameter deformation using $\xi_0=1+\sigma$ 
and $\xi_9=-\sigma$ with $\sigma\in(-1,\infty)$. The resulting distribution is slightly oblate for $-1<\sigma<0$ and prolate to any degree for $\sigma>0$.  

Our definition (\ref{H-def}) can be extended by including terms with higher order products of the scalar products $(\hat n_3\cdot\vec v)$ and $(\hat n_3\cdot\vec v)$. 
The only restriction we impose, besides positivity, is the condition $H_\xi(\vec v) = H_\xi(-\vec v)$. This means there are no terms where the sum of the exponents is odd,  like for example $(\hat n_3\cdot\vec v)(\hat n_1\cdot\vec v)^2$. 
The reason we require that $H_\xi(\vec v)$ is even under the transformation $\vec v \to -\vec v$ is as follows. 
We said above that when general distributions are considered the polarization tensor is obtained from equations (\ref{pi-even}, \ref{pi-odd}) by making the replacement $n(k) \to n(\vec k) = C_\xi \, n\big(k H_\xi(\vec v)\big)$. In fact, there is an extra contribution to the anisotropic version of equation (\ref{pi-even}) of the form
\bea 
\Pi^{ij}_{\rm extra}(p_0,\vec p) \sim g^2\int\frac{d^3 k}{(2\pi)^3} \big(n_R(\vec k)+\bar n_R(\vec k)\big)
\left(\frac{v^i v^j}{p_0 - \vec p\cdot \vec v +i\epsilon}-\frac{v^i v^j}{p_0 + \vec p\cdot \vec v +i\epsilon} \right)\, \label{pi-extra} 
\eea
that integrates to zero for any distribution that is even under the transformation $\vec k \to -\vec k$.  
If the condition $H_\xi(\vec v) = H_\xi(-\vec v)$ is not satisfied, equation (\ref{pi-extra}) would produce a non-zero contribution to the polarization tensor that dominates over the HL terms, and is not present in the result obtained from semiclassical kinetic theory (which is equivalent to the HL expression).
 
We will calculate the anisotropic polarization tensor using equations (\ref{pi-even-2}, \ref{pi-odd-2}) with the replacement $n(k) \to C_\xi \,n\left(k H_\xi(\vec v)\right)$ and similarly for $\bar n(k)$. 
We define 
$\tilde k = k H_\xi(\vec v)$ 
and perform a straightforward change of variable so that the integral over $\vec k$ is written as the product of an integral over the new variable $\tilde k$ and a factored integral involving the angular variables. 
Equations (\ref{pi-even-2}, \ref{pi-odd-2}) become
\bea
&& \Pi^{ij}_{{R\rm even}}(p_0,\vec p) =   (m_D^2)_R \,  C_\xi\,\int\frac{d\Omega}{4\pi} \frac{1}{H_\xi^4(\vec v)} v^i\,{\cal M}^l(\Omega)\,    \left(\delta^{jl} + \frac{p^l v^j}{P\cdot V + i\epsilon}\right) \label{pi-even-3}\\
&& \Pi^{ij}_{R{\rm odd}}(p_0,\vec p) =   i \frac{g^2 \mu_R}{2\pi^2} \epsilon^{ijm} p  \,C_\xi\, \int \frac{d\Omega}{4\pi} \frac{1}{H_\xi^3(\vec v)} {\cal M}^l(\Omega)
 \left(p_0 \delta^{lm} + \frac{(p_0 v^m-p^m)p^l}{P\cdot V + i\epsilon}
\right) ~~~~~~~~~~~ \label{pi-odd-3}
\label{PIanio}
\eea
where we have defined
\bea
{\cal M}^l(\Omega) = \left(\frac{1}{2k} \frac{\partial (k^2 H_\xi^2(\vec v))}{\partial k^l}\right)
\label{calM-def}
\eea
and used $(m^2_D)_R = g^2(T^2/3+\mu_R^2/\pi^2)$. 
The full expression for $\Pi^{ij}$ is obtained using equation (\ref{combo}), as in the isotropic case.

We close this section by explaining how we define the function of the anisotropy parameters which we call $C_\xi$, which can be thought of as a normalization of the anisotropic distribution function. 
This factor is determined by enforcing the condition
\bea
\int \frac{d^3k}{(2\pi)^3} \frac{n_{\mu=0}(k)}{k} \equiv C_\xi\int \frac{d^3k}{(2\pi)^3}\frac{1}{k} \, n_{\mu=0}\big(k H_\xi(\vec v)\big)
\eea
from which we obtain
\bea
C_\xi^{-1} = \int\frac{d\Omega}{4\pi}\frac{1}{H^2_\xi(\vec v)}\,. \label{norm-def}
\eea
This choice of normalization gives that when the chemical potentials are zero the mass parameter in equation (\ref{def-mD}) is independent of the anisotropy parameters. 
At zero chemical potential the entire spectrum of collective modes depends on this one scale (since we have assumed massless partons - see under equation (\ref{nn-def})), and it is therefore natural to adopt a normalization that leaves the Debye mass invariant under a change of the anisotropy parameters. 
We also note that using equations (\ref{pi-even-3}, \ref{calM-def}, \ref{norm-def}) it is easy to see that the even part of the polarization tensor is unchanged when the complete set of anisotropy parameter are scaled uniformly: $\xi_i \to \Lambda \xi_i$ where $\Lambda$ is any positive constant. This means that when we compare the spectrum of collective modes obtained from different choices of anisotropy parameters, we are comparing the effects of a specific deformation, and not the influence of an overall change of scale. 

\subsection{Projection operators}
 
To decompose our general expression for the $3\times 3$ polarization tensor we need a complete set 
of nine projection operators. This can be done most conveniently using three normalized unit vectors that are related to the plasmon momentum $\vec p$ and the two anisotropy vectors $\hat n_1$ and $\hat n_3$. 
From these three-vectors we construct three ortho-normal vectors which we call $(\hat p, ~n_f,~m_F)$. We define
\bea
&& \hat p = \frac{\vec p}{p} \nn\\
&& n_f = \frac{\tilde n_f}{\sqrt{\tilde n_f\cdot \tilde n_f}} \text{~~with~~} \tilde n_f = n_3 - (n_3\cdot \hat p) \hat p\,~~ \nn\\
&& m_F=\frac{\tilde m_F}{\sqrt{\tilde m_F\cdot \tilde m_F}} \text{~~with~~} \tilde m_F = \tilde m_f - (n_f\cdot\tilde m_f)\,n_f \text{~~and~~} \tilde m_f = n_1 - (n_1 \cdot \hat p) \hat p\,.
\eea
Using these unit vectors the projection operators are defined as
\bea
&& P_1^{\ij} = m_F^i m_F^j \,,~~ P_2^{\ij} = \hat p^i \hat p^j \,,~~ P_3^{\ij} = n_f^i n_f^j \nn\\
&& P_4^{\ij} = \hat p^i n_f^j + n_f^i \hat p^j \,,~~ P_5^{\ij} = \hat p^i m_F^j + m_F^i \hat p^j \,,~~ P_6^{\ij} =  n_f^i m_F^j + m_F^i n_f^j \nn\\
&& P_7^{\ij} =  n_f^i \hat p^j - \hat p^i n_f^j \,,~~P_8^{\ij} =  n_f^i m_F^j -  m_F^i n_f^j \,,~~P_9^{\ij} =  \hat p^i m_F^j - m_F^i \hat p^j \,.
\label{proj-all}
\eea 
They satisfy conditions of the form $P^{ij}_m P^{ji}_n = {\cal T}_{mn}$ where the 9$\times$9 tensor ${\cal T}$ is
\bea
\left(
\begin{array}{ccccccccc}
 2 P_1 & 0 & 0 & 0 & P_5 & P_6 & 0 & P_8 & P_9 \\
 0 & 2 P_2 & 0 & P_4 & P_5 & 0 & P_7 & 0 & P_9 \\
 0 & 0 & 2 P_3 & P_4 & 0 & P_6 & P_7 & P_8 & 0 \\
 0 & P_4 & P_4 & 2 \left(P_2+P_3\right) & P_6 & P_5 & 0 & P_9 & P_8 \\
 P_5 & P_5 & 0 & P_6 & 2 \left(P_1+P_2\right) & P_4 & P_8 & P_7 & 0 \\
 P_6 & 0 & P_6 & P_5 & P_4 & 2 \left(P_1+P_3\right) & -P_9 & 0 & -P_7 \\
 0 & P_7 & P_7 & 0 & P_8 & -P_9 & -2 \left(P_2+P_3\right) & -P_5 & P_6 \\
 P_8 & 0 & P_8 & P_9 & P_7 & 0 & -P_5 & -2 \left(P_1+P_3\right) & -P_4 \\
 P_9 & P_9 & 0 & P_8 & 0 & -P_7 & P_6 & -P_4 & -2 \left(P_1+P_2\right) \\
\end{array}
\right)\,.\nonumber\\
\label{proj-table}
\eea
The projectors in equation (\ref{proj-all}) form a complete basis, and therefore the polarization tensor can be decomposed as
\bea
\Pi^{ij} = \pi_1 P_1^{ij}+\pi_2 P_2^{ij}+\pi_3 P_3^{ij}+\pi_4 P_4^{ij}+\pi_5 P_5^{ij}+\pi_6 P_6^{ij}+i \pi_7 P_7^{ij}+i \pi_8 P_8^{ij}+ i \pi_9 P_9^{ij}\,
\label{finalPi}
\eea
where we have omitted the functional arguments to shorten the notation. 
A decomposition that is related to the first six of our projection operators was defined in Ref. \cite{Ghosh:2020sng}. 
The last three projection operators are anti-symmetric in their indices, and the corresponding dressing functions can only be non-zero if the chiral chemical potential is non-zero. 
We invert the Dyson equation (\ref{dyson-def}) using (\ref{proj-table}) and obtain the propagator
\bea
{\cal D} D^{ij} &=& \left[ \left(p_0^2-\pi _2\right) \left(P^2-\pi _3\right)-\pi _4^2-\pi _7^2 \right] P_1^{ij} 
+ \left[ \left(P^2-\pi _1\right) \left(P^2-\pi _3\right)-\pi _6^2-\pi _8^2 \right] P_2^{ij} \nn \\
&+& \left[ \left(p_0^2-\pi _2\right) \left(P^2-\pi _1\right)-\pi _5^2-\pi _9^2 \right] P_3^{ij}  
+ \left[ \left(P^2-\pi _1\right)\pi _4 +\pi _5 \pi _6+\pi _8 \pi _9 \right] P_4^{ij}   \nn \\
&+& \left[\left(P^2-\pi _3\right) \pi _5 +\pi _4 \pi _6+\pi _7 \pi _8  \right] P_5^{ij} 
+ \left[ \left(p_0^2-\pi _2\right) \pi _6 +\pi _4 \pi _5-\pi _7 \pi _9 \right] P_6^{ij}   \nn\\
&+& i \left[ \left(P^2-\pi _1\right) \pi _7 +\pi _5 \pi _8-\pi _6 \pi _9  \right] P_7^{ij} 
+ i \left[ \left(p_0^2-\pi _2\right) \pi _8 +\pi _5 \pi _7+\pi _4 \pi _9  \right] P_8^{ij} \nn \\
&+& i \left[ \left(P^2-\pi _3\right)  \pi _9 -\pi _6 \pi _7+\pi _4 \pi _8 \right] P_9^{ij} \label{prop9}
\eea
where we have defined
\bea
{\cal D} &=& \left(p_0^2-\pi _2\right) \left(P^2-\pi _1\right) \left(P^2-\pi _3\right)
-\left(P^2-\pi _1\right) \left(\pi _4^2+\pi _7^2\right) 
-\left(p_0^2-\pi _2\right) \left(\pi_6^2+\pi _8^2\right)  \nonumber \\
&& - \left(P^2-\pi _3\right) \left(\pi _5^2+\pi _9^2\right) 
 -2\left(\pi _4 \pi _5 \pi _6-\pi _7 \pi _9 \pi _6+\pi _5 \pi _7 \pi _8+\pi _4 \pi _8\pi _9\right)\,.\label{den9}
\eea   
The anisotropic dispersion relations are the solutions of the dispersion equation, ${\cal D}=0$. 

\subsection{Coordinate system}
\label{coord}
To make further process we must define a  coordinate sytem. The vector along the beam axis can be chosen as $\hat n_3' = (0,0,1)$ and the vector that defines the direction of asymmetry in the transverse plane can then be chosen as
$\hat n_1' = (1,0,0)$. The external momentum vector then has the general form 
$\vec p^{\;\prime} = p(\sin\theta \cos\phi, \sin\theta \sin\phi, \cos\theta)$.
We note that using these coordinates it is easy to see that the parametrization in equation (\ref{H-def}) is equivalent to an expansion in spherical harmonics $Y_{lm}(\theta,\phi)$ with $l\in(0,4)$ and $m\in(-l,l)$.  

We can define coordinates that are more convenient for calculational purposes by performing a rotation \cite{Kasmaei:2018yrr}. 
We rotate counter-clockwise about the $z$-axis with angle $\theta_z = \pi/2-\phi$ and then counter-clockwise about the $x$-axis with $\theta_x=\theta$. 
The result is that the basis vectors become
  \bea
&& \hat n_1 =  (\sin(\phi), \cos(\theta)\cos(\phi), \sin(\theta)\cos(\phi)) \nn\\
&& \hat n_3 = (0, -\sin(\theta), \cos(\theta)) \nn\\
&& \hat p = \vec p /p= (0,0,1) \,.\label{new-coords}
  \eea
Any scalar or pseudo-scalar quantity will give the same result when calculated with either the original or the rotated coordinate system.   
This can be easily be seen from the fact that the scalar products $(\hat p\cdot\hat n_1)$, $(\hat p\cdot\hat n_3)$ and $(\hat n_1\cdot\hat n_3)$, and the pseudo-scalar $\epsilon^{ijk}\hat n_1^i \hat n_3^j \hat p^k$, are invariant. Likewise the dispersion relations are the same in either coordinate system. 
Another useful thing about the coordinates in equation (\ref{new-coords}) is that the isotropic limit has a simple form. 
It is easy to show using equation (\ref{new-coords}) that
$\delta^{ij}-\hat p^i\hat p^j = P_1^{ij}+P_3^{ij}$, $\hat p^i\hat p^j = P_2^{ij}$ and $\epsilon^{ijm}p^m = P^{ij}_8$. 
In the isotropic limit there are only three different non-zero components in equation (\ref{finalPi}) which are 
$\pi_1=\pi_3=\Pi_T$,
$\pi_2=\Pi_L$ and 
$\pi_8 = \Pi_A$. 
Using these results one can see that equations (\ref{prop9}, \ref{den9}) reduce to (\ref{prop-aka}). 

The variables in the angular integrations in equations (\ref{pi-even-3}, \ref{pi-odd-3}) are independent variables, and will be written in component form as 
\bea
&& \vec v = (\sin(\theta')\cos(\phi'),\sin(\theta')\sin(\phi'),\cos(\theta'))\,. \label{new-v}
\eea
We will use the notation $x=\cos(\theta)$ and $x'=\cos(\theta')$.

\subsection{Anisotropy parameters}
\label{sec-YI}

The anisotropy parameters $\xi_9$ and $\xi_2$ produce uniform stretching or squeezing of the isotropic distribution along the beam axis (in the case of $\xi_9$), and along the vector $\hat n_1$ in the transverse plane (for $\xi_2$). 
We have introduced additional anisotropy parameters that can be used to produce distribution functions that are deformed relative to the isotropic distribution in more general ways, and produce dispersion relations with more structure. 
As discussed in section \ref{Hdef-section}, we consider only distributions that satisfy the condition $f(\vec k) = f(-\vec k)$. 
This restricts the form of the terms in equation (\ref{H-def}). 
Each term is chosen so that $e_1+e_3$ is even, where the exponent of the factor $(\vec n_1\cdot\vec v)$ is denoted $e_1$ and the exponent of the factor $(\vec n_3\cdot\vec v)$ is called $e_3$. 
There are two different possibilities: $e_1$ and $e_3$ individually even, or $e_1$ and $e_3$ individually odd.

\begin{figure}[htb]
\includegraphics[scale=0.29]{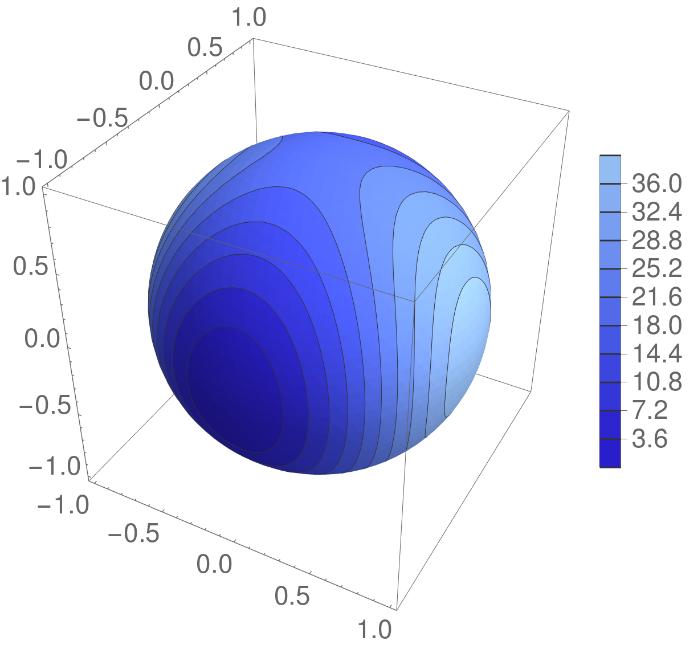}
\includegraphics[scale=0.32]{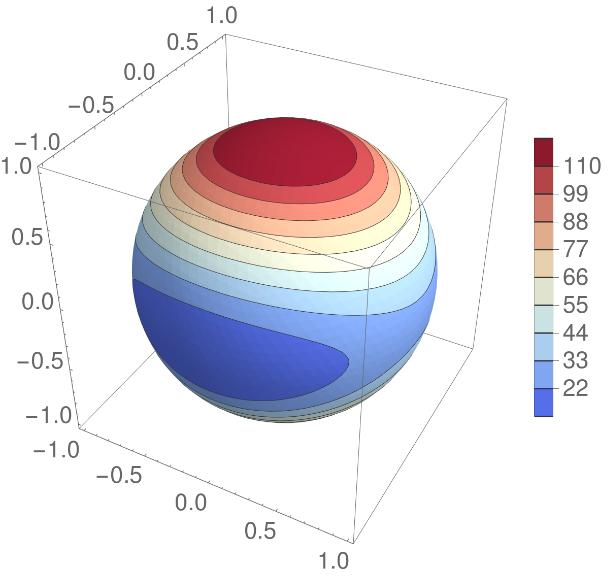}
\caption{Spherical plots of the coefficients $\xi_2(\Omega)$ (left panel) and $\xi_9(\Omega)$ (right panel) that correspond to the mapping of distribution 6 in table \ref{table-choices} onto an elliptically anisotropic distribution. The values in the figure are obtained with $\beta=1$ and $\mu=0$. \label{bailey-fig}}
\end{figure}

The elliptically anisotropic distribution considered in \cite{Kasmaei:2018yrr} depends only on the two parameters $\xi_9$ and $\xi_2$, both of which have $e_1$ and $e_3$ individually even.
Any distribution for which $e_1$ and $e_3$ are individually even for all terms in (\ref{H-def}) can be mapped onto an elliptically anisotropic distribution with angularly dependent parameters. 
%
We consider for example
the distribution constructed with the set of anisotropy parameters $\xi_0=1$, $\xi_4=20$, $\xi_9=10$, $\xi_{11}=20$, $\xi_{14}=60$
(distribution 6 in table \ref{table-choices}).
This distribution is equivalent to an elliptically anisotropic distribution with effective coefficients $\xi_2(\Omega)$ and $\xi_9(\Omega)$ 
as shown in figure \ref{bailey-fig}. 
The new parameters $\xi_4$, $\xi_{11}$ and $\xi_{14}$ effectively modify the constant value of $\xi_9$ and generate a non-zero $\xi_2$.

The parameters $\xi_6$, $\xi_8$ and $\xi_{13}$ are coefficients of terms in equation (\ref{H-def}) with $e_1$ and $e_3$ individually odd. 
These terms produce a distortion of the isotropic distribution that is qualitatively different. 
Some examples of the dispersion relations obtained from distributions with non-zero values of these coefficients are presented in section \ref{sec-dispersion}.

\begin{figure}[H]
\includegraphics[scale=0.38]{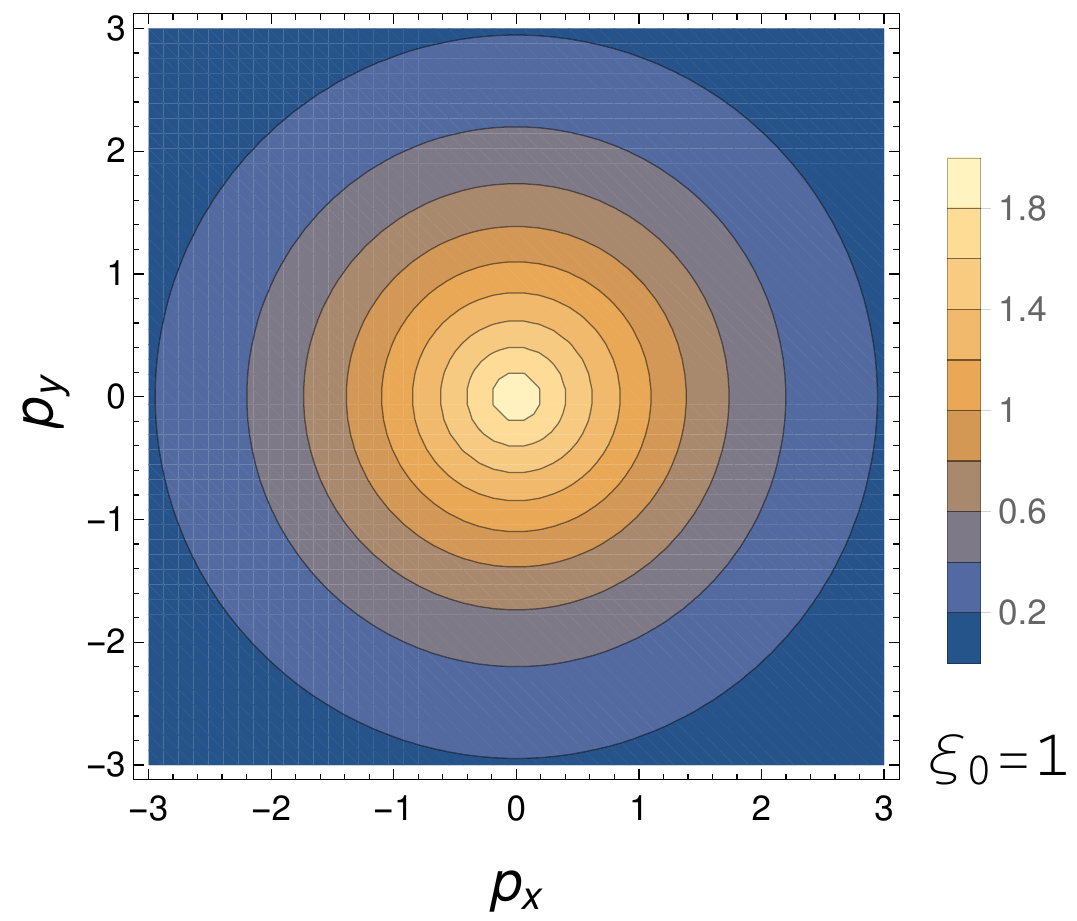}
\includegraphics[scale=0.38]{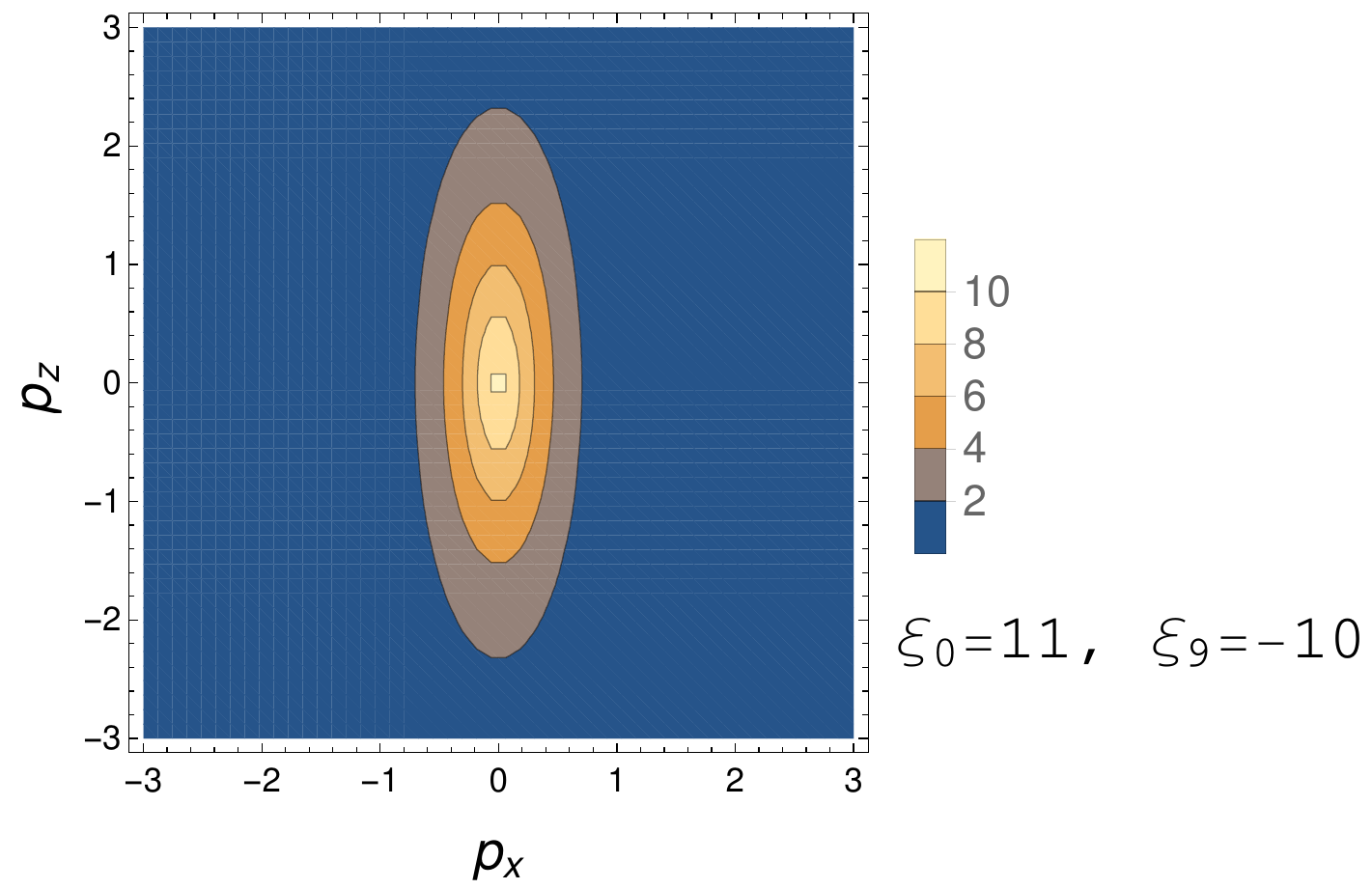} 
\includegraphics[scale=0.38]{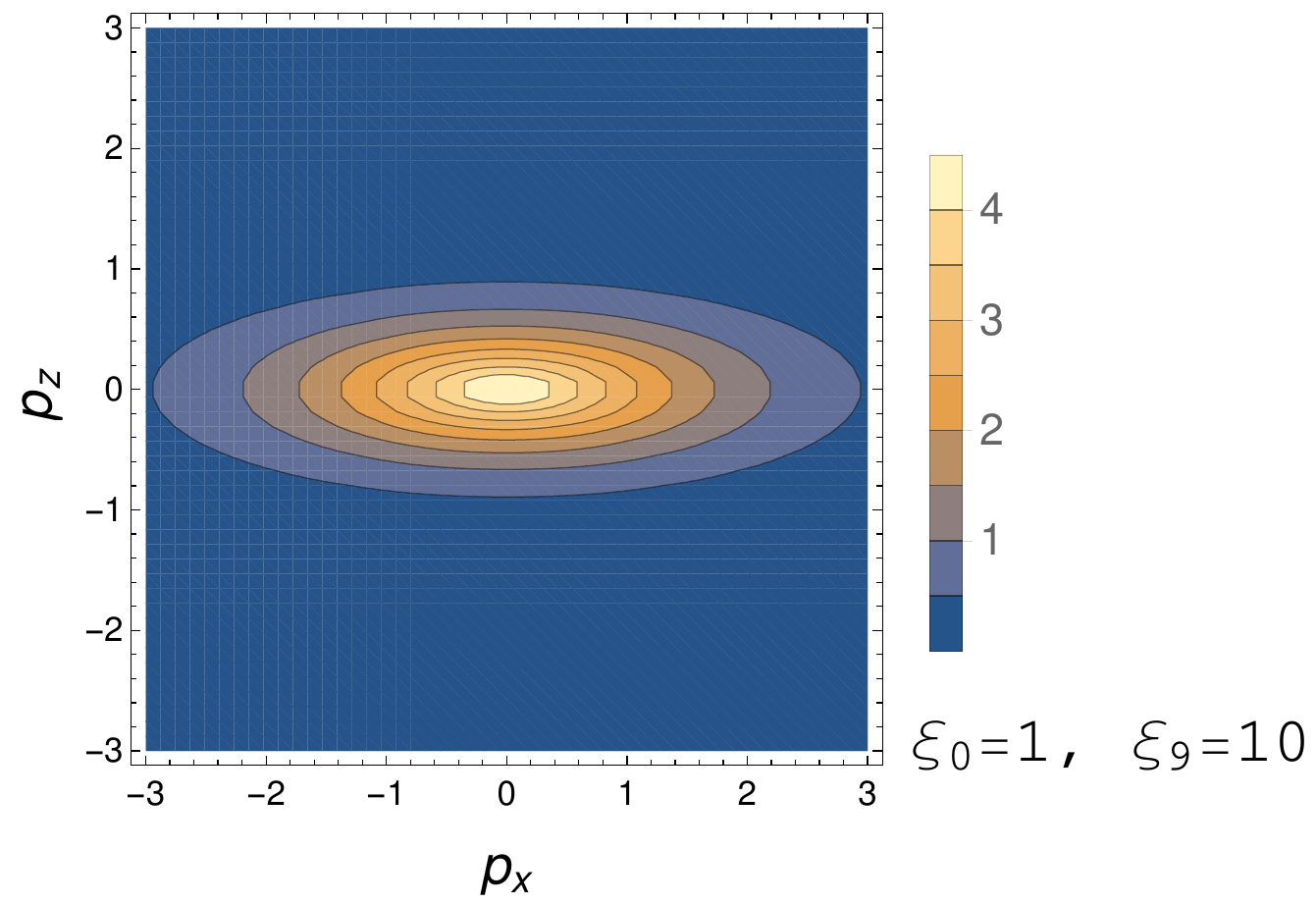} 
\includegraphics[scale=0.35]{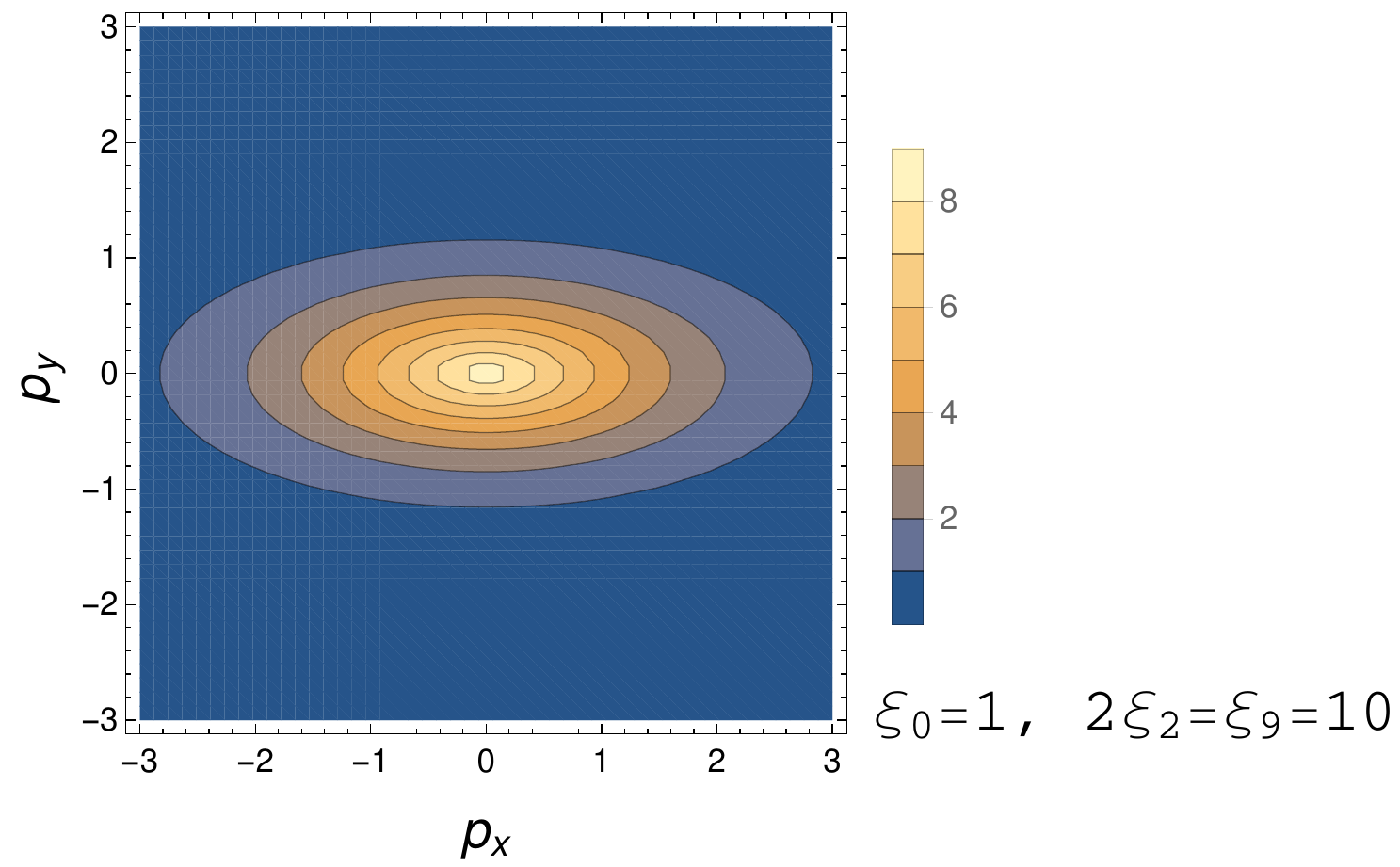} 
\includegraphics[scale=0.35]{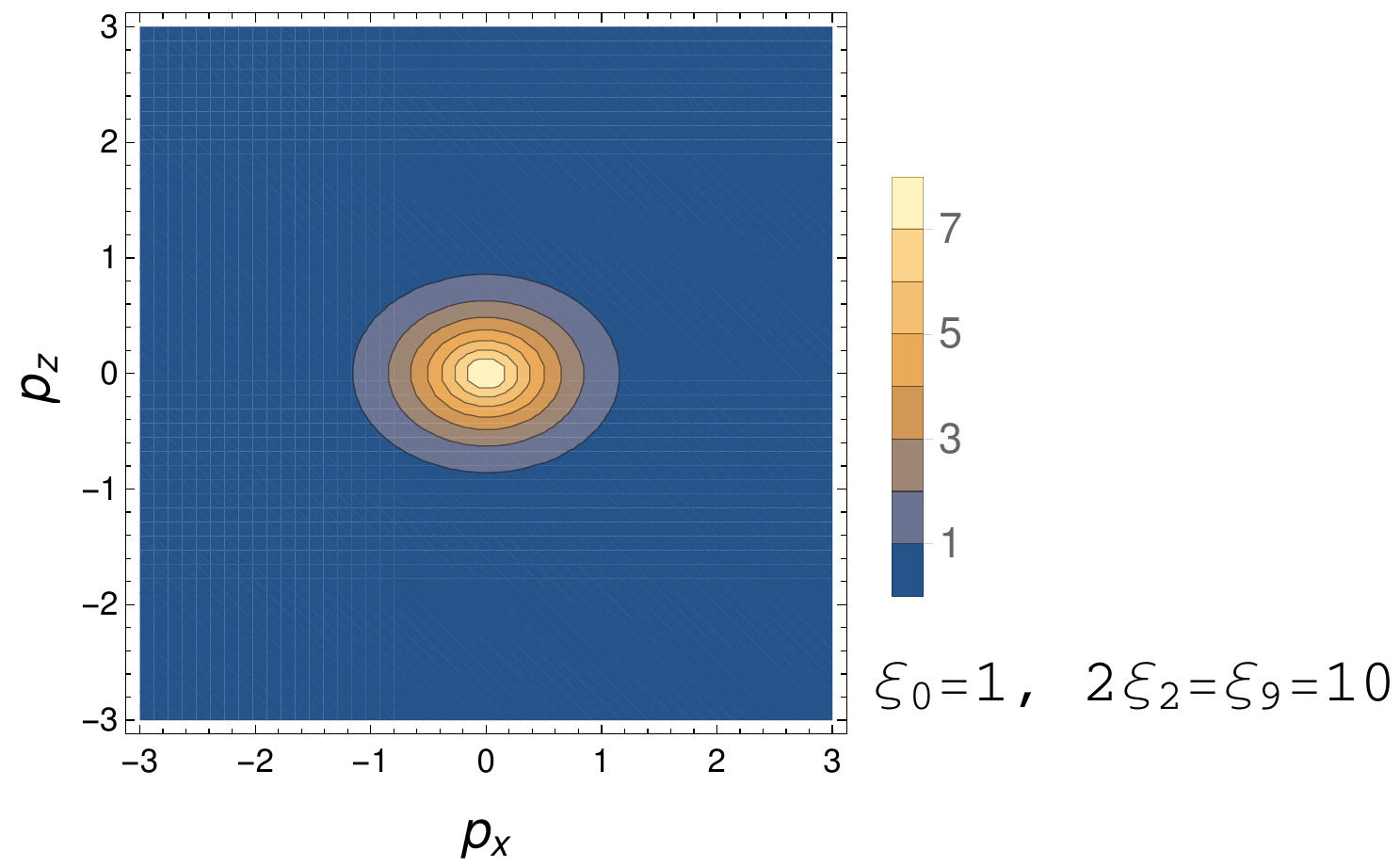} 
\includegraphics[scale=0.35]{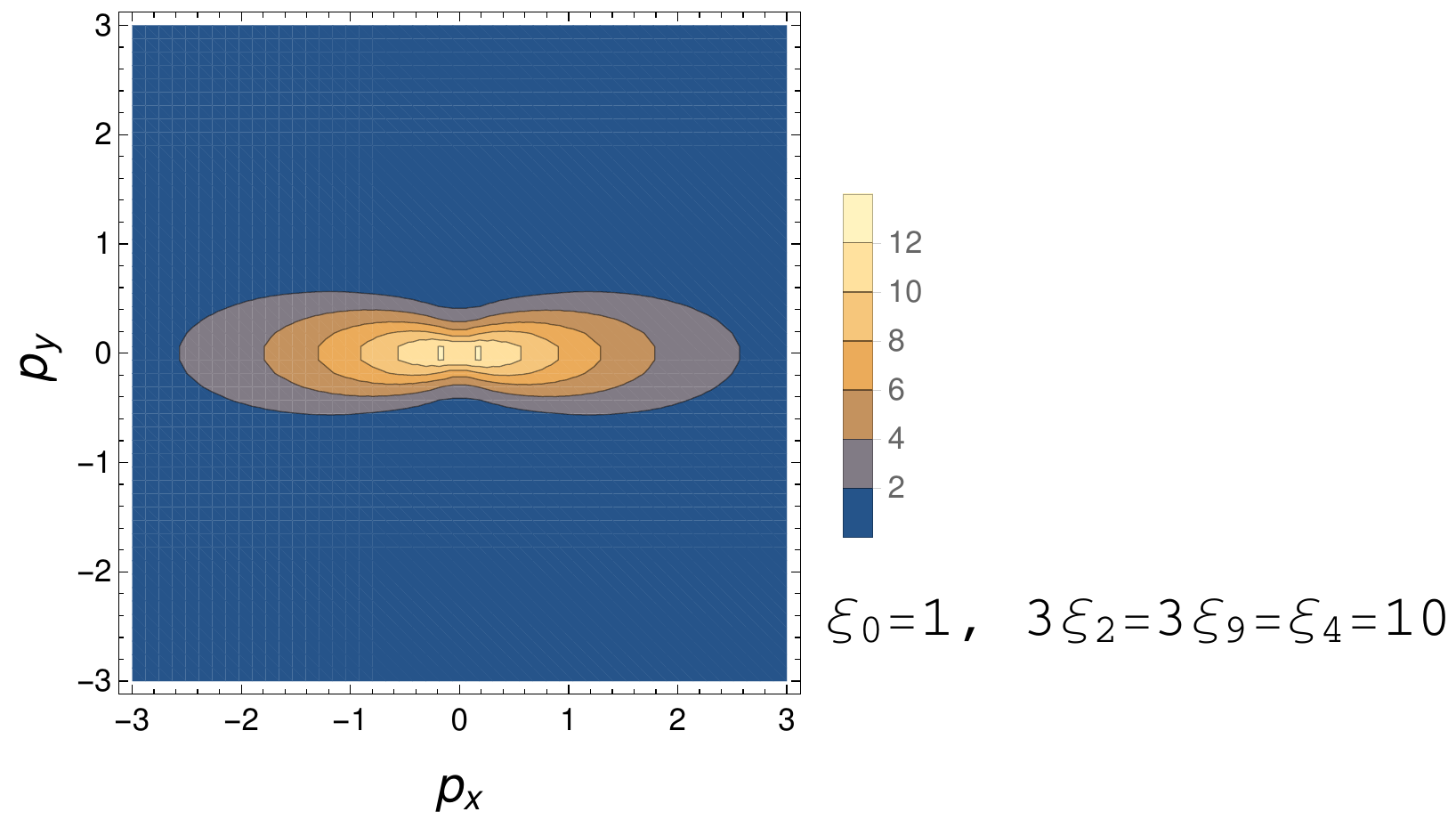} 
\caption{Contour plots of the distribution for different choices of the anisotropy parameters. In each case the parameters that are not specified are set to zero. \label{contours}}
\end{figure}

To illustrate more directly the roles of the different anisotropy parameters we look at  plots of  $C_\xi \, n\big(k H_\xi(\vec v)\big)$ with $\beta=1$ and $\mu=0$, in the $(p_x,p_y)$ and $(p_x,p_z)$ planes. 
In figure \ref{contours} we show the results  for several choices of anisotropy parameters. 
The first panel shows, for reference, the isotropic distribution in the $(p_x,p_y)$ plane, the figure for the $(p_x,p_z)$ plane is identical. In the second panel we show a prolate distribution in the $(p_x,p_z)$ plane, and one sees clearly that the distribution is stretched along the beam axis. 
The third panel shows the squeezing of an oblate distribution along the beam axis. 
In the fourth and fifth panels we show an elliptically oblate distribution, which is asymmetric in both planes. 
In the sixth panel we show an example of the kind of structure that can be obtained with a specific choice of the extra anisotropy parameters we have introduced. 

\section{Dispersions relations}
\label{sec-dispersion}
\subsection{Analytic structure}

The isotropic and anisotropic dressing functions have the same  analytic structure, which can be determined from equations (\ref{pi-even-2}, \ref{pi-odd-2}), using that equations (\ref{new-coords}, \ref{new-v}) give $P\cdot V = p(\hat p_0-x')$ where $\hat p_0 = p_0/p$. 
When $\hat p_0$ is pure real the epsilon prescription is needed to define the integrands at $\hat p_0=x'$. 
Shifting the integration variable $\vec v\to -\vec v$ one shows that $\pi_i^*(\hat p_0) = \pi_i(-\hat p_0)$, where the subscript $i$ indicates any component of the polarization tensor. This means that when $\hat p_0$ is real, the real part of each dressing function is even in $\hat p_0$ and the imaginary part is odd. We also note that all dressing functions are pure real for $\hat p_0>1$, since the imaginary part comes from the discontinuity at $\hat p_0=x'$ and $|x'|<1$. 
When $\hat p_0$ is not pure real the epsilon prescription is not needed, and we have $\pi_i^*(\hat p_0) = \pi_i(\hat p^*_0)$. When $p_0$ is pure imaginary $\pi_i(\hat p^*_0) = \pi_i(\hat p_0)$, which can be shown by shifting the integration variable, and therefore the dressing functions are pure real and even in $\hat p_0$. 
The information above can be summarized as
\bea
{\rm Re}(\hat p_0)>1 \text{~and~} {\rm Im}(\hat p_0)=0 ~~ &\to& ~~ \pi_i(\hat p_0) \text{~ real ~and~even} \nn \\
\hat p_0 \text{~imaginary~}  ~~ &\to& ~~ \pi_i(\hat p_0) \text{~ real ~and~even} \nn \\[3mm]
{\rm Re}(\hat p_0)<1 \text{~and~} {\rm Im}(\hat p_0)=0 ~~&\to&~~ 
\left\{ \begin{array}{ll} 
{\rm Re}\big( \pi_i(\hat p_0)\big) \text{~~ even} \\
{\rm Im} \big(\pi_i(\hat p_0)\big) \text{~~ odd\,.} \end{array}\right.		\nn 
\eea

\subsection{Method}

The integrand for the polarization tensor is given by equations (\ref{pi-even-3}, \ref{pi-odd-3}, \ref{calM-def}). 
The nine scalar components $\pi_1$-$\pi_9$ are obtained by contracting with the appropriate projection operators, which are easily obtained from equations (\ref{proj-table}, \ref{finalPi}). These calculations are straightforward but tedious, and they are therefore done with {\it Mathematica}. 
The resulting integrals are calculated numerically using Gauss-Legendre quadrature. 
For real valued $\hat p_0$, the real part of the dressing functions has a single integrable singularity at $x'=\hat p_0$ that is easily handled with a numerical principal part prescription, and the imaginary part can be obtained from the residue of the pole. For imaginary valued $\hat p_0$ there are no difficulties. After the dressing functions are obtained, the dispersion equation is solved with a standard binary search algorithm. 

\subsection{Real solutions}

\begin{figure}[htb]
\centering
\includegraphics[scale=0.73]{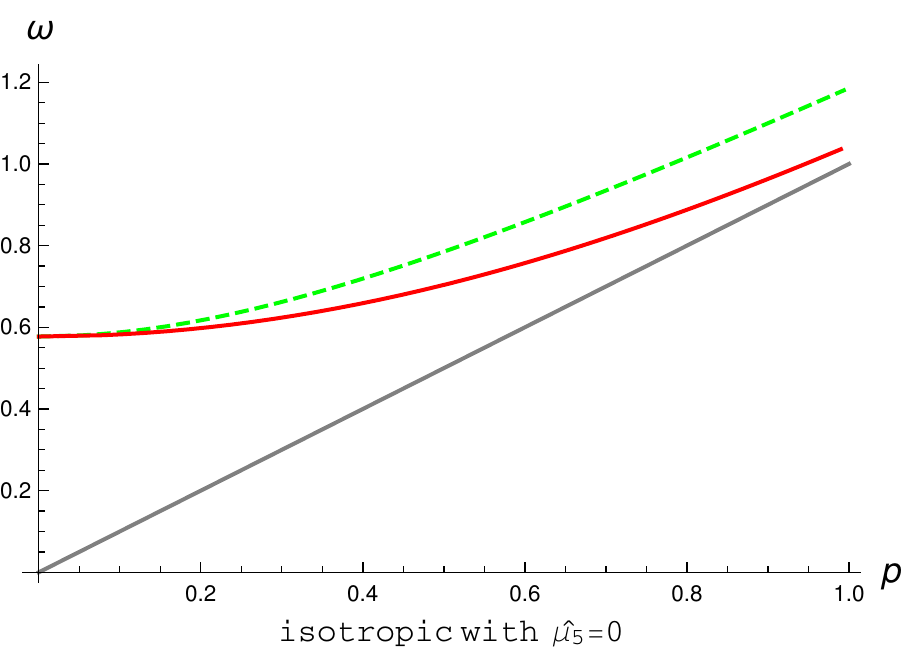}
\includegraphics[scale=0.73]{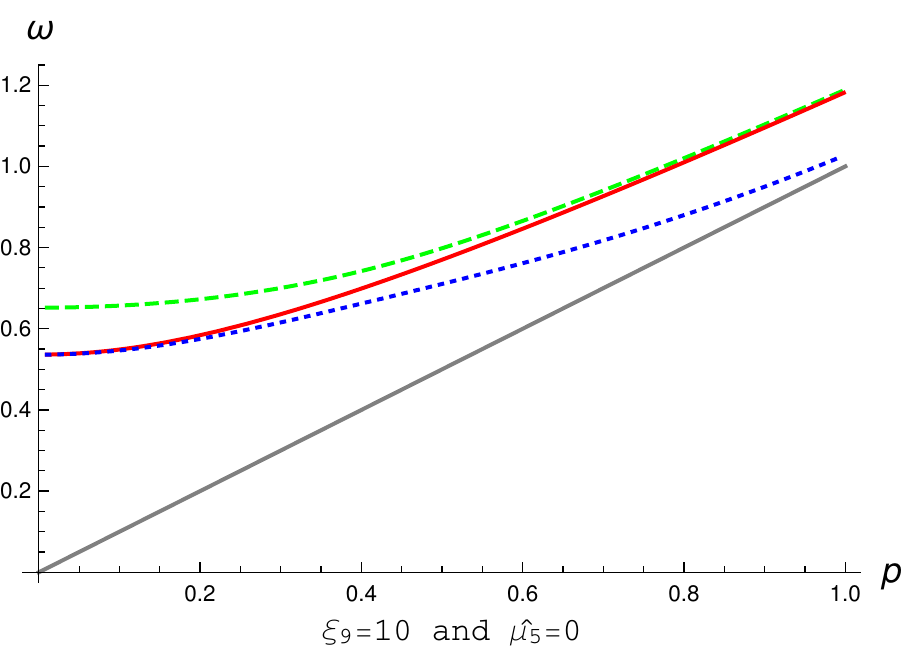}
\includegraphics[scale=0.73]{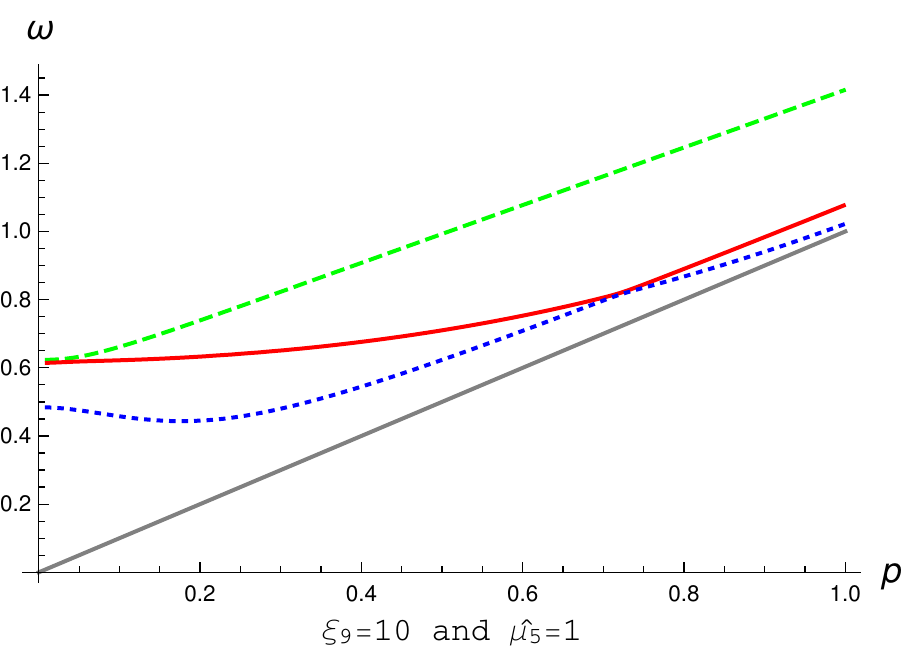}
\includegraphics[scale=0.73]{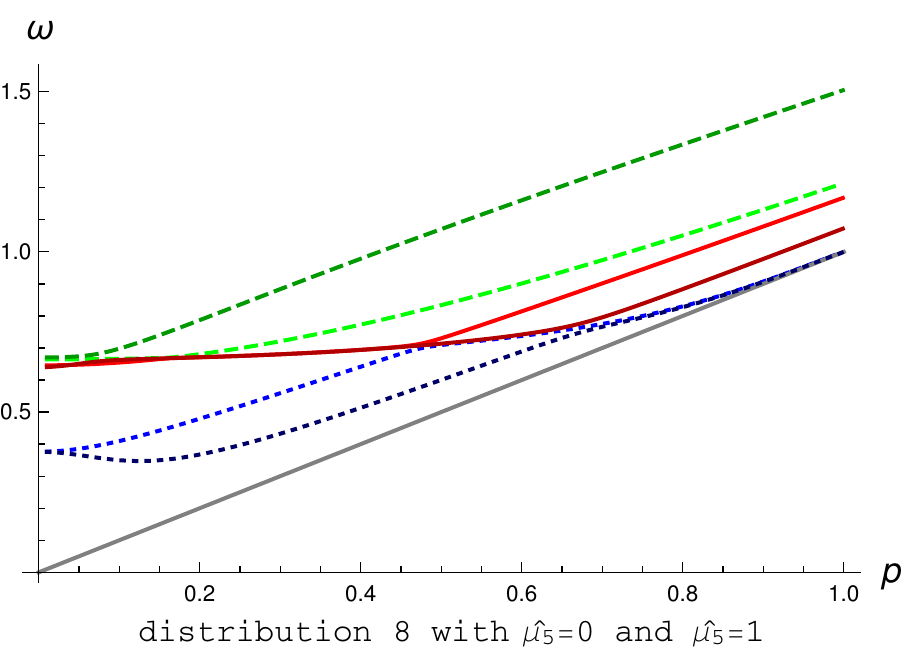}
\caption{Real solutions for distributions 1, 3 and 8 (as defined in table \ref{table-choices}). In the bottom right panel the lighter lines are the soutions with $\hat\mu_5=0$ and the dark lines are $\hat\mu_5=1$. In all graphs the light gray line shows the light cone.\label{realsolutions}}
\end{figure}

For any choice of the anisotropy parameters and the chiral chemical potential $\mu_5$ there are three real solutions. 
We show some examples of the dispersion relations in figure \ref{realsolutions}.

\subsection{Imaginary solutions}
We are primarily interested in imaginary solutions, because of their potential to influence plasma dynamics. 
We will compare the imaginary modes produced by different choices of the anisotropy parameters, and explore their dependence on these parameters as a function of wave vector. 
We have previously used the notation $\gamma(\vec p)$ to indicate the dependence of the dispersion relation on the wave vector, but in this section we will use the more explicit notation $\gamma(p,x,\phi)$ where $x=\cos(\theta)$. 
For different sets of anisotropy parameters
we compare the maximum value of the imaginary solution,
the position of this maximum, 
and the integral of the solution over its domain, which we write  $\int\gamma\equiv \int_0^\infty dp \int_{-1}^1 dx\int_0^{2\pi} d\phi \,\gamma(p,x,\phi)$.

We define three variables that characterise the degree of oblateness, the transverse eccentricity, and the azimuthal asymmetry of a given distribution:
\bea
&& \chi = \frac{\langle k_x^2\rangle + \langle k_y^2\rangle}{2\langle k_z^2\rangle}-1 \nn \\
&& \varepsilon = \frac{\langle k_y^2\rangle-\langle k_x^2\rangle}{\langle k_y^2\rangle + \langle k_x^2\rangle} \nn \\
&& v_2 = \frac{\int dk \, k^2 \int d\phi \, \cos(2\phi) \, n\big(k H_\xi(\vec v)\big)_{x=0}}{\int dk \, k^2 \int d\phi \, n\big(k H_\xi(\vec v)\big)_{x=0}} \,.\label{params}
\eea
The angle brackets denote averaging over the momentum phase space, for example $\langle k_x^2\rangle = \int d^3 k k_x^2 C_\xi n(k H_\xi(\vec v))$.

In table \ref{table-choices} we show some of the main features of the largest imaginary solution of the dispersion relation for several different distributions. 
In all cases the anisotropy parameters that are not specified are zero.
The first distribution is the isotropic one. The second and third have spheroidal anisotropy, with parameters that make them prolate and oblate, respectively. The fourth is a distribution with spheroidal anisotropy that is strongly oblate. 
The fifth is an elliptically anisotropic distribution. 
Distributions 6-9 are examples constructed using the additional anisotropy parameters that we have introduced.  
Distributions 6 and 7 involve only terms for which $e_1$ and $e_3$ are even (see the discussion at the beginning of section \ref{sec-YI}). 
The second to last column shows that the maximum of the largest mode can be much greater than the result obtained from a distribution with spheroidal anisotropy with a comparable value of the oblateness parameter $\chi$. 
In addition, the last column shows that the integral of the solution over its full domain is greatly enhanced, relative to the result obtained from a comparable distribution with spheroid oblateness. 
For example, distribution 7 has approximately the same value of $\chi$ as distribution 3, but $\gamma_{\rm max}$ is about twice as big, and the integral of the solution is almost 9 times larger. 
Distributions 8 and 9 are examples for which $e_1$ and $e_3$ are individually odd for some of the terms in (\ref{H-def}). 
The magnitudes of the modes produced by these distributions are similar to what is obtained from distributions without these odd coefficients, but the angular dependence of the solutions is much more complex. At the end of this section we  show several plots to demonstrate this. 
Finally, the table shows that the effect of a non-zero chiral chemical potential is always to increase the magnitude and domain of the imaginary solution. The solutions produced with even moderate anisotropy are much larger than those obtained with the same chemical potential in an isotropic system. 
\begin{table}[H]
\centering 
\begin{tabular}{|l|l|c | c | c|c|c|c|} 
\hline                   
 \# &   ~~ $\xi_i$ ~~ & ~$\hat\mu_5$ ~ &~~ $\chi$ ~~& ~~ $\varepsilon$ ~~ & ~~ $v_2$  ~~ & ~~ $\gamma_{\rm max}(p,x,\phi)$ ~~ & ~~  $\int \gamma$ ~~ \\ [0.5ex]
\hline\hline
 1&  $\xi_0=1$ & 0 & 0 & 0 & 0 & - & - \\
  & & $0.3$ &  &  &  & $4.7\times10^{-3}$(.19,-,-) & 0.017 \\
\hline
2 & $(\xi_0,\xi_9)=(11,-10)$ & 0  & -0.91 & 0 & 0 & 0.16(0.45,0,-) & 0.59 \\
  &                          & 0.3   &       &   &   & 0.23(0.68,0,-) & 0.86 \\
 \hline
3 &$(\xi_0,\xi_9)=(1,10)$ & 0  & 10 & 0 & 0 & 0.19(0.63,1,-) & 0.08 \\
 &  & 0.3  &  &  &  & 0.34(1.07,1,-) & 0.24 \\
  \hline
4 & $(\xi_0,\xi_9)=(1,500)$ & 0  & 500 & 0 & 0 & 0.39(1.57,1,-) & 0.32 \\
 &  & 0.3  &  &  &  & 1.60(21.0,1,-) & 6.90 \\
  \hline
5 &  $(\xi_0,\xi_2,\xi_9)=(1,5,10)$ & 0 & 5.42 & 0.71 & 0.59 & 0.20(0.58,1,0) & 0.17 \\
  & & 0.3 &  &  &  & 0.29(0.88,1,0) & 0.34 \\
\hline
6 & $\xi_{0,4,9,11, 14}=(1,20,10,20,60)$  & 0 & 17.9 & 0.85 & 0.7 & 0.33(0.78,1,1.49) & 0.56 \\
 & & 0.3 &  &  &  & 0.50(1.23,1,0.06) & 1.15 \\
\hline
7 & $\xi_{0,4,9,11, 14}=(1,40,-2,40,55)$  & 0 & 10.1 & 0.91 & 0.85 & 0.32(0.68,0.79,0) & 0.70 \\
 & & 0.3 &  &  &  & 0.43(1.06,0.80,0) & 1.27 \\
\hline
8 & $\xi_{0,2,4,6,8,9,11,13,14}=$ & 0 & 8.15 & 0.90 & 0.97 & 0.36(0.88,0.67,$\pi$) & 0.61 \\
&  $(1,8,37,-21,21,8,15,-28,40)$  & 0.3 &  &  &  & 0.56(1.65,0.63,$\pi$) & 1.27 \\
\hline
9 & $\xi_{0,9,13,14}=(1,1,20,20)$ & 0 & 1.80 & -0.35 & 0 & 0.23(0.70,0.85,0) & 0.16 \\
 &  & 0.3 &  &  &  & 0.31(0.94,0.86,0) & 0.27 \\
\hline
\end{tabular}
\caption{Different sets of anisotropy parameters, the corresponding values of the parameters $\xi$, $\varepsilon$ and $v_2$ defined in equation (\ref{params}), the size and location of the largest imaginary solution, and the integral of the solution over its domain. In each case the anisotropy parameters that are not specified are set to zero. }
\label{table-choices}
\end{table}

For a given distribution, the number of imaginary solutions is 0 or 1 or 2, depending on the magnitude and direction of the wave vector. 
For each mode there is a critical vector $p_{\rm crit}(x,\phi,\xi_i)$ such that the imaginary mode appears for $p$ less than this critical value.  
Analytic results for the critical wave vectors can be obtained in the limit of weak anisotropy using a Nyquist analysis. The method is explained in detail in Appendix \ref{nyquist-sec}, where we show how to identify the angular variables that produce the largest critical wave vectors for a given set of anisotropy parameters. 
To understand the structure of the solutions obtained from a given distribution in more detail, we show several figures below. 

In figure \ref{fig-aa} we compare the imaginary solutions obtained from the distributions 3 and 6 for a specific choice of angles. The figure illustrates the concept of the critical wave vector. 
For example, in the right panel of figure \ref{fig-aa}, the critical wave vectors for modes 1 and 2 
are, respectively, 2.33 and 0.62. 
It is true in general that when the domain of the solution increases, the maximum value of the solution also increases, as seen in figure \ref{fig-aa}.  
\begin{figure}[H]
\centering
\includegraphics[scale=.5]{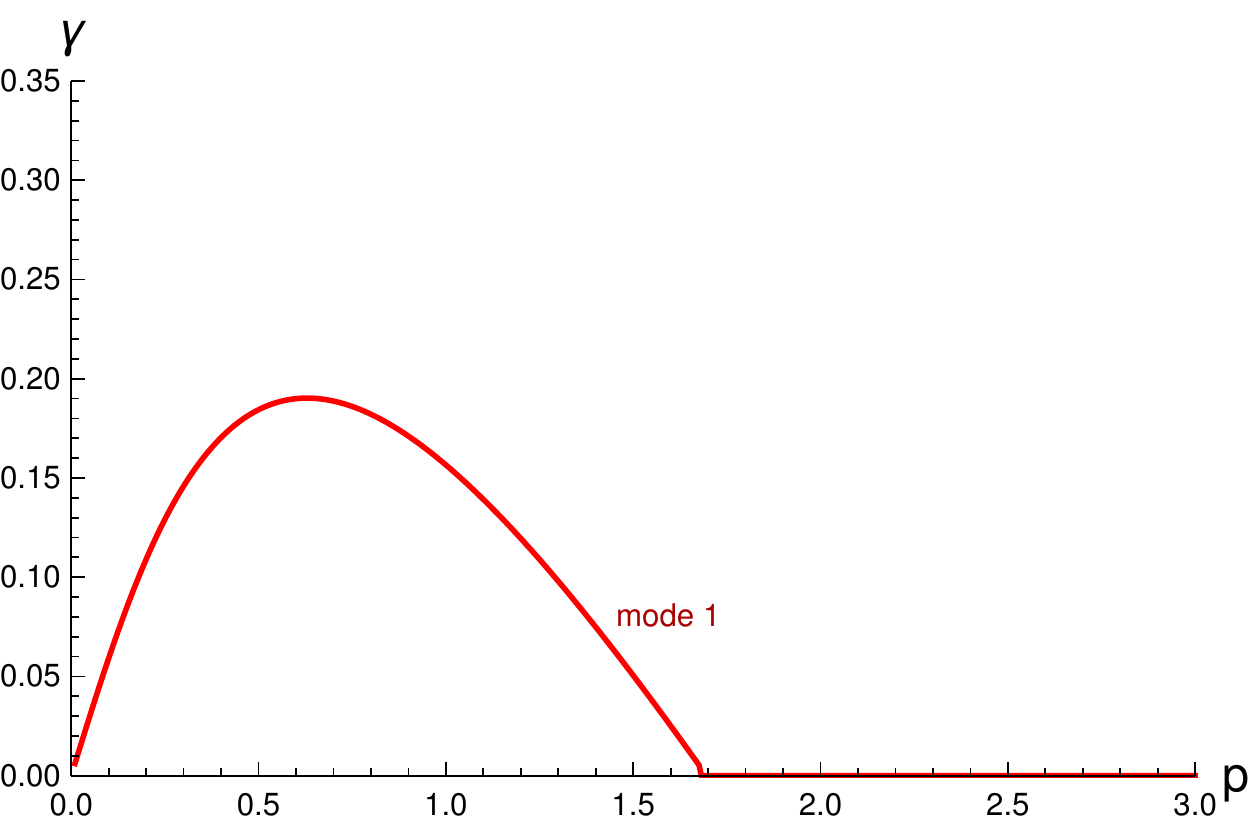}
\includegraphics[scale=.5]{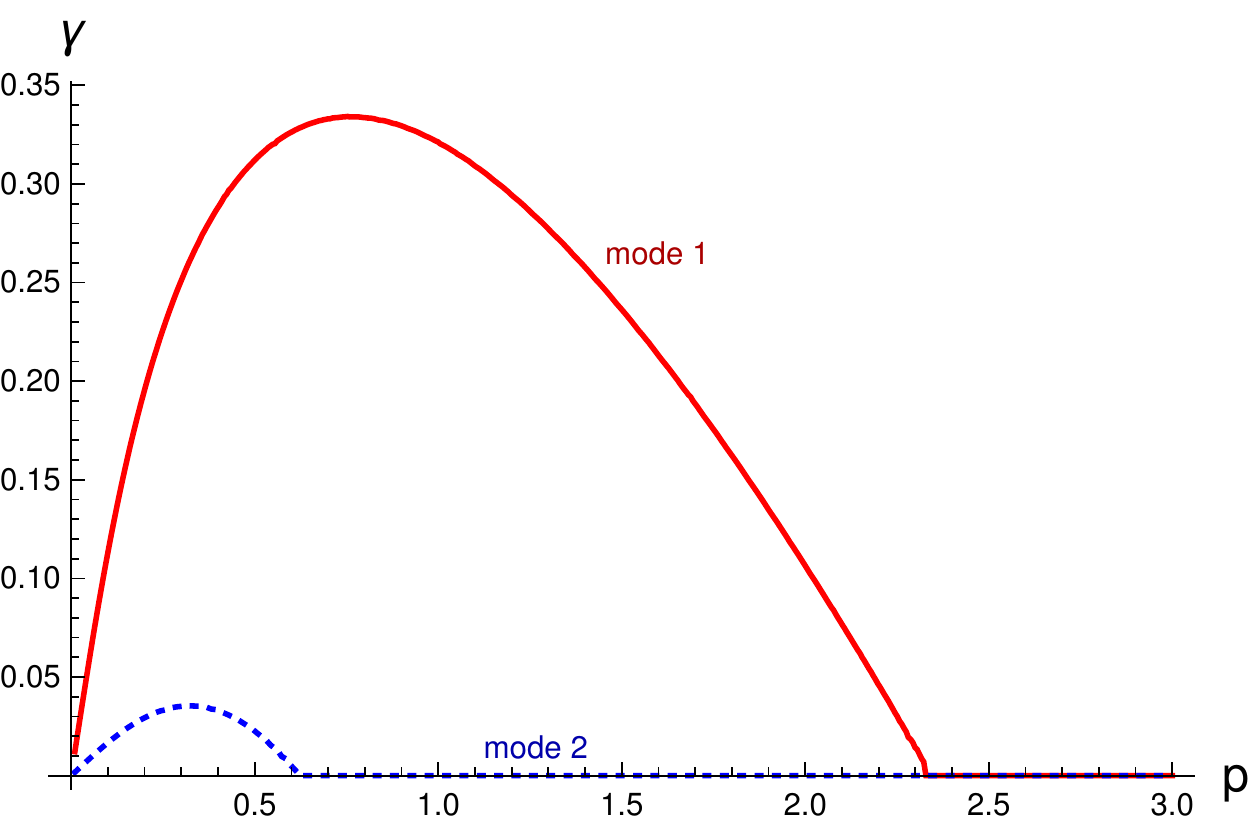} 
\caption{The imaginary solutions of the dispersion equation ${\cal D}=0$ obtained from distributions 3 (left panel) with $\hat\mu_5=0$ and $x=1$,  and the two imaginary solutions obtained from distribution 6 (right panel) with $\hat\mu_5=0$, $x=1$ and $\phi=\pi/2$. \label{fig-aa}}
\end{figure}

%

\begin{figure}[H]
\centering
\includegraphics[scale=.95]{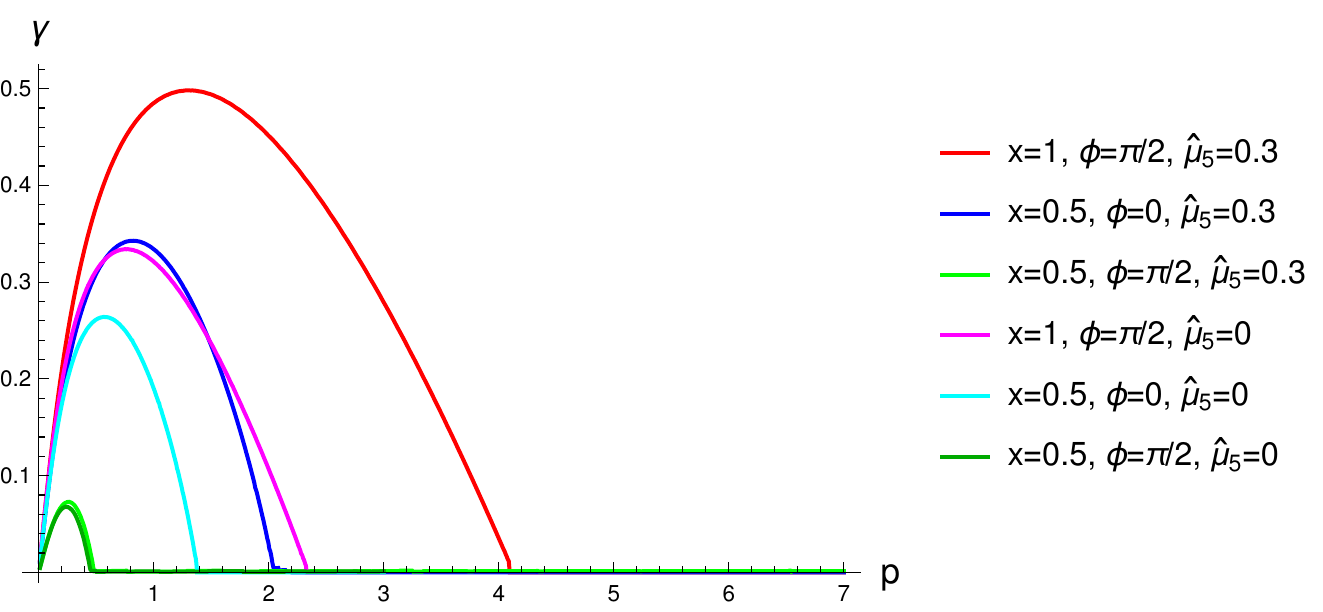} 
\caption{The largest imaginary solution of the dispersion equation ${\cal D}=0$ using the distribution 6, two different chiral chemical potentials, and several different values of the angular variables. \label{fig-bb}}
\end{figure}

In figure \ref{fig-bb} we show the largest imaginary mode obtained from distribution 6 for different choices of angles, and two different values of the chiral chemical potential. 
The figure shows that increasing the chiral chemical potential increases the domain and size of the solution, but not uniformly as a function of the angular variables. 

\begin{figure}[H]
\centering
\includegraphics[scale=.53]{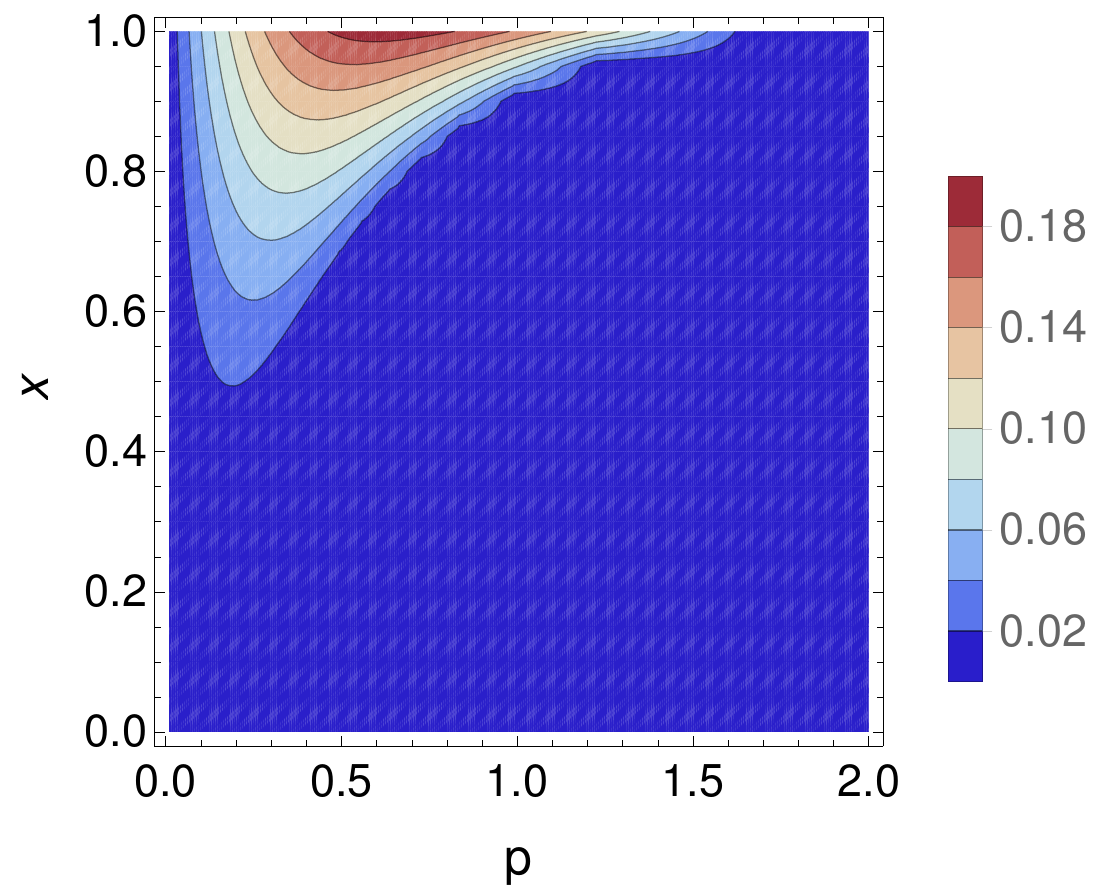} 
\includegraphics[scale=.53]{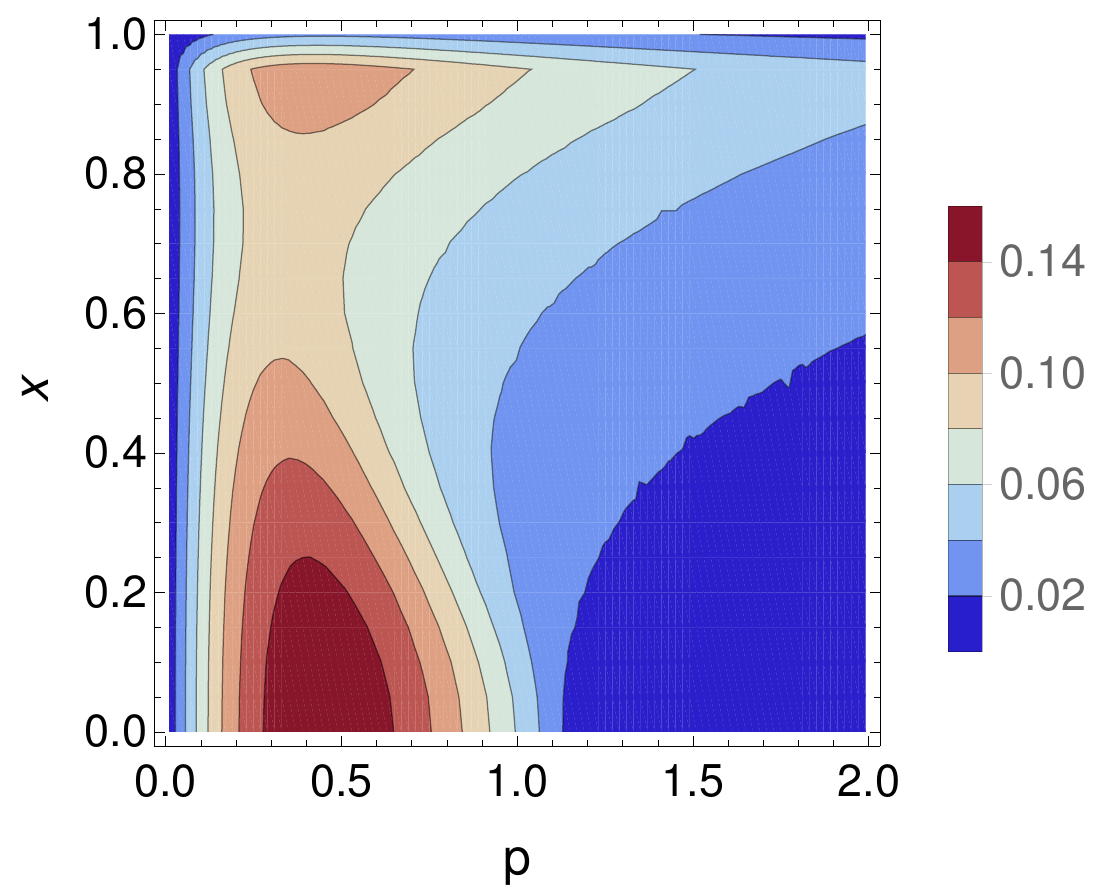}
\includegraphics[scale=.41]{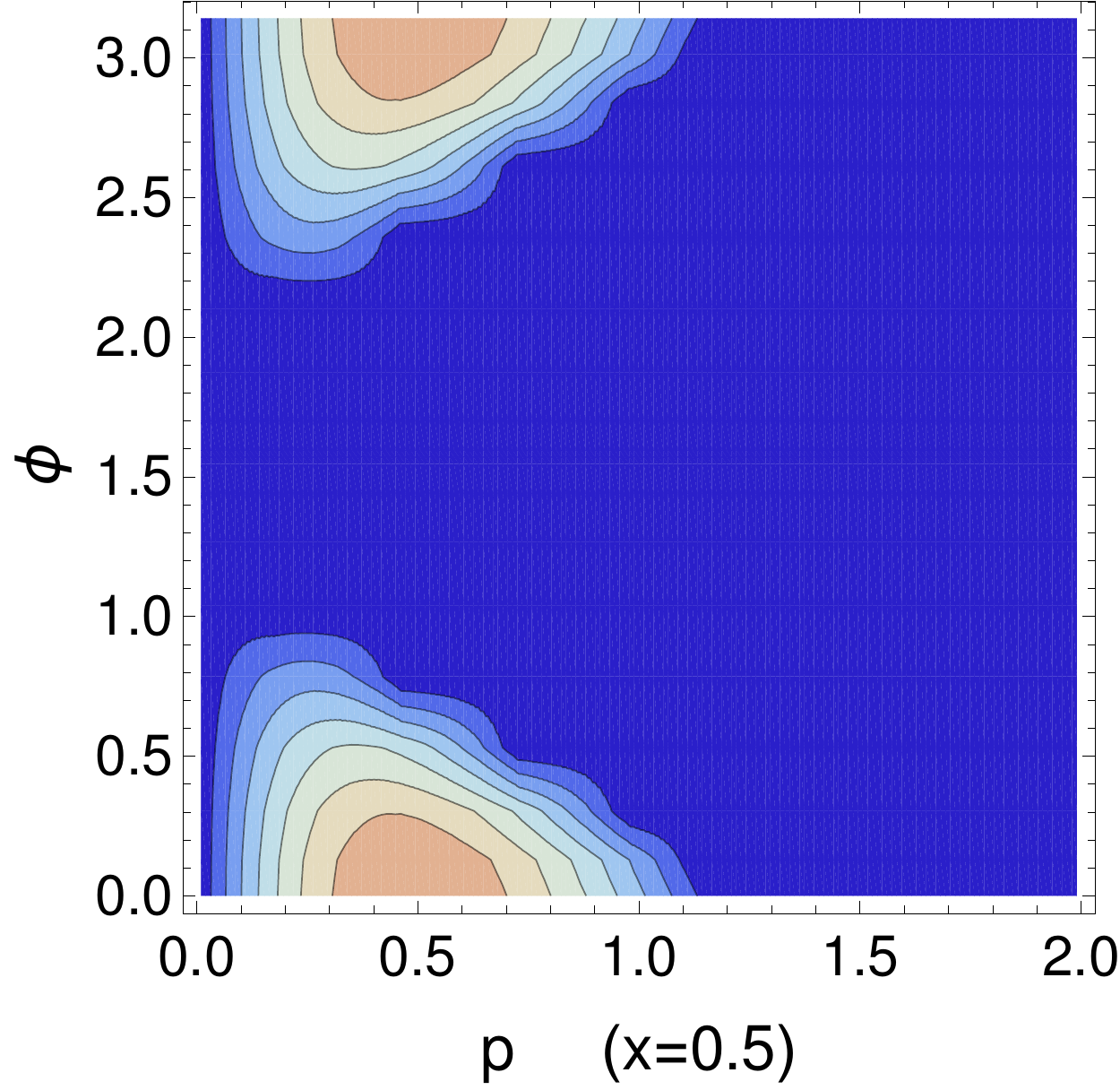}
\includegraphics[scale=.58]{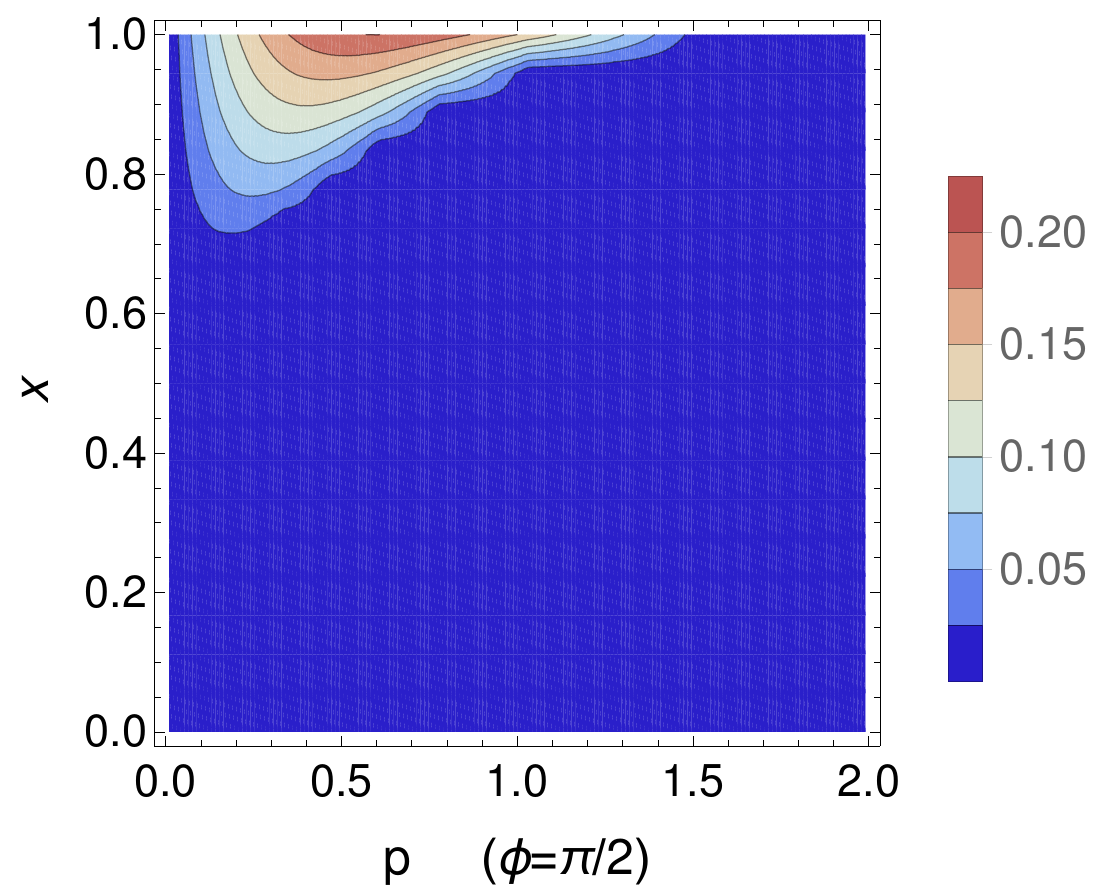}
\caption{Contour plots of the largest imaginary solution for several distributions with $\hat\mu_5=0$. The top left and top right panels show the $\phi$ independent solution obtained from distributions 3 and 2, respectively. The bottom panels are solutions obtained from distribution 5.\label{fig-anio3_10-aa}}
\end{figure}

The dependence of the imaginary modes on the angular variables can be seen most clearly using contour plots. 
In the top left panel of figure \ref{fig-anio3_10-aa} we show a contour plot of the magnitude of the largest imaginary solution for distribution 3 with $\hat\mu_5=0$, as a function of $p$ and $x$ (there is no $\phi$ dependence). The solution is non-zero in a small region of the phase space where $x$ is close to 1 and $p$ is small. 
In the top right panel of figure \ref{fig-anio3_10-aa} we show the largest imaginary solution for distribution 2 with $\hat\mu_5=0$,
which is also $\phi$ independent. The magnitude of the solution is smaller, but the domain is larger. 
The bottom two panels show the $\phi$ dependent solutions obtained from distribution 5. 

\begin{figure}[htb]
\centering
\includegraphics[scale=.275]{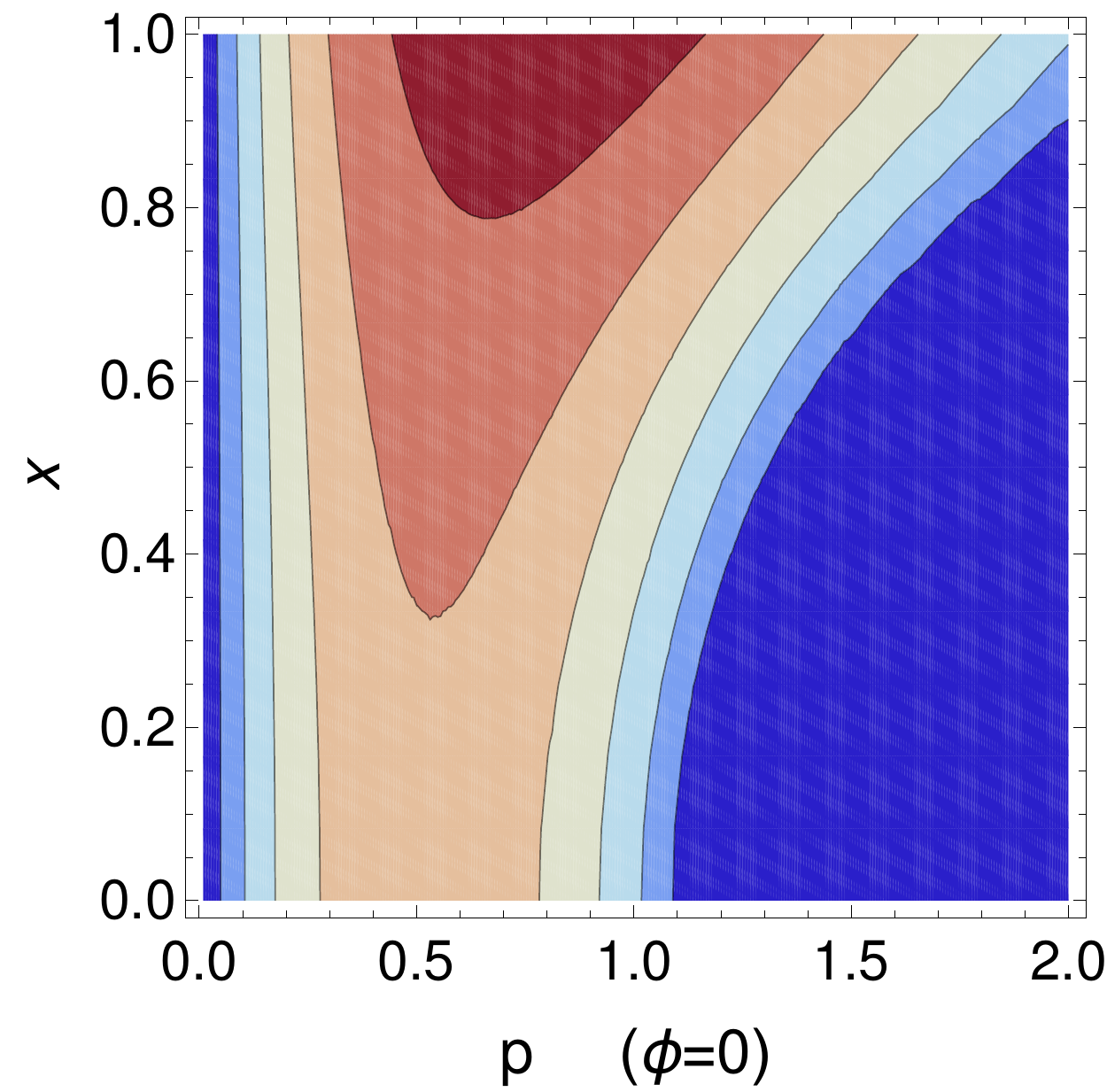} \quad
\includegraphics[scale=.275]{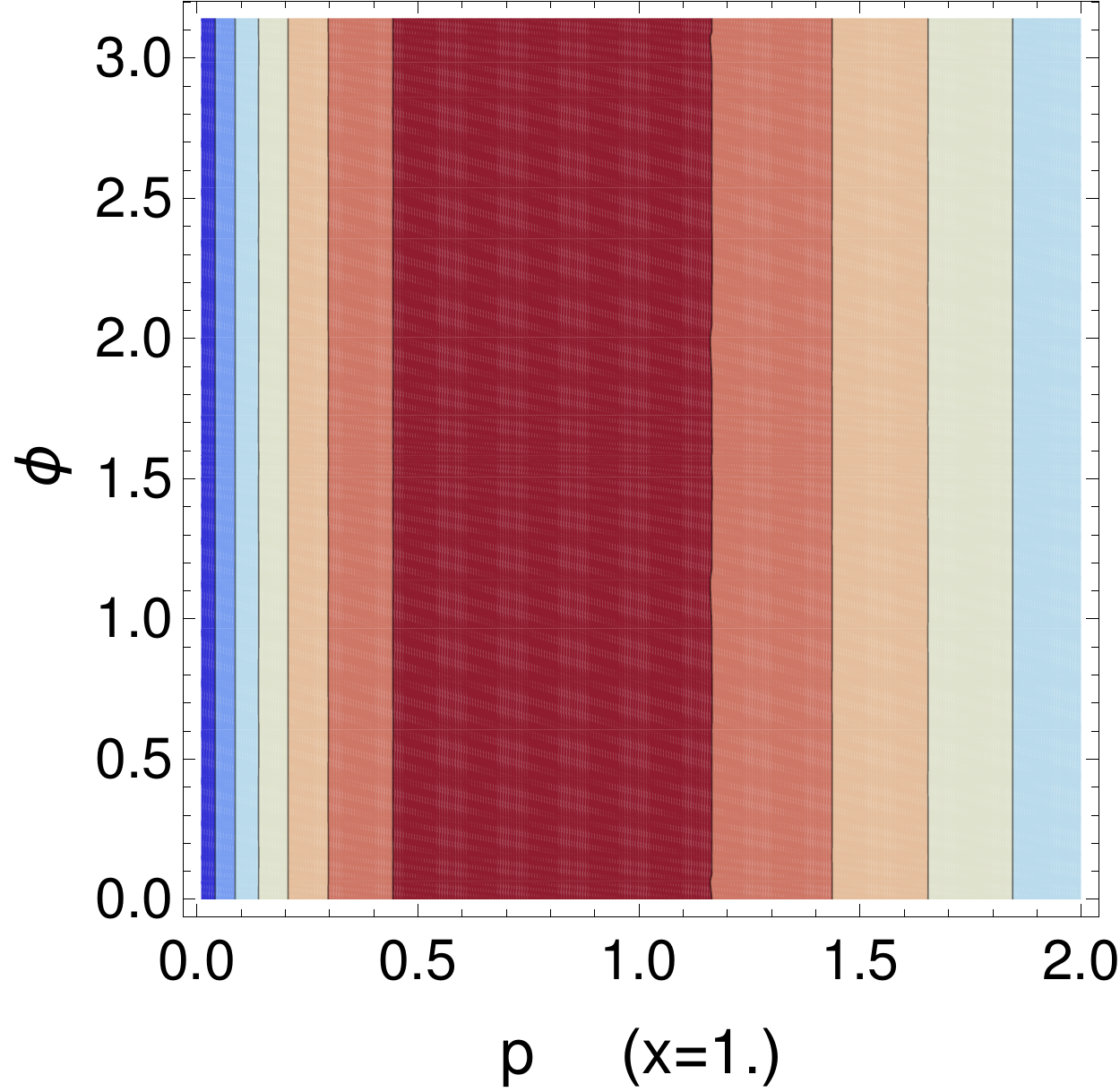} \quad
\includegraphics[scale=.275]{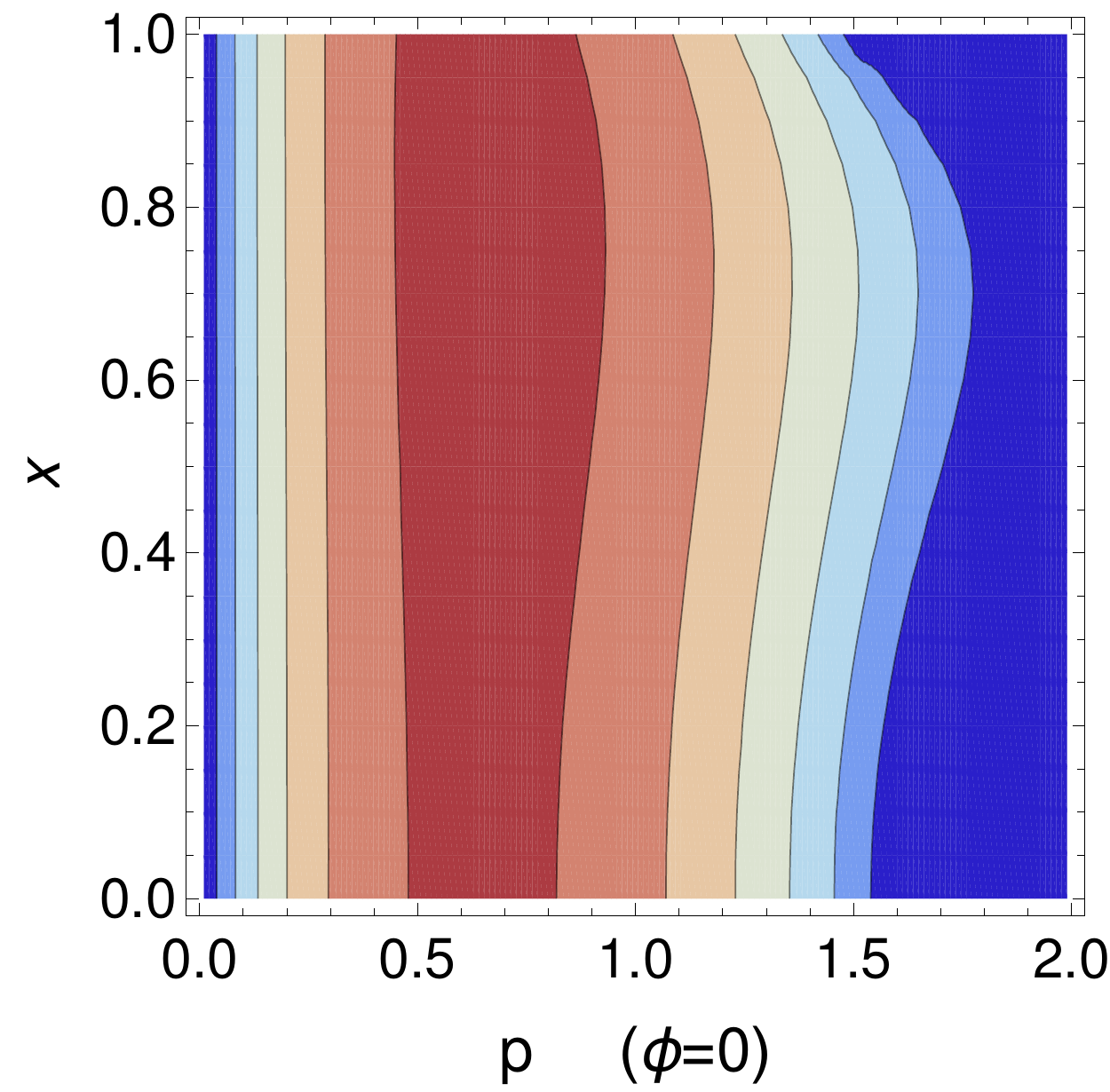} \quad
\includegraphics[scale=.38]{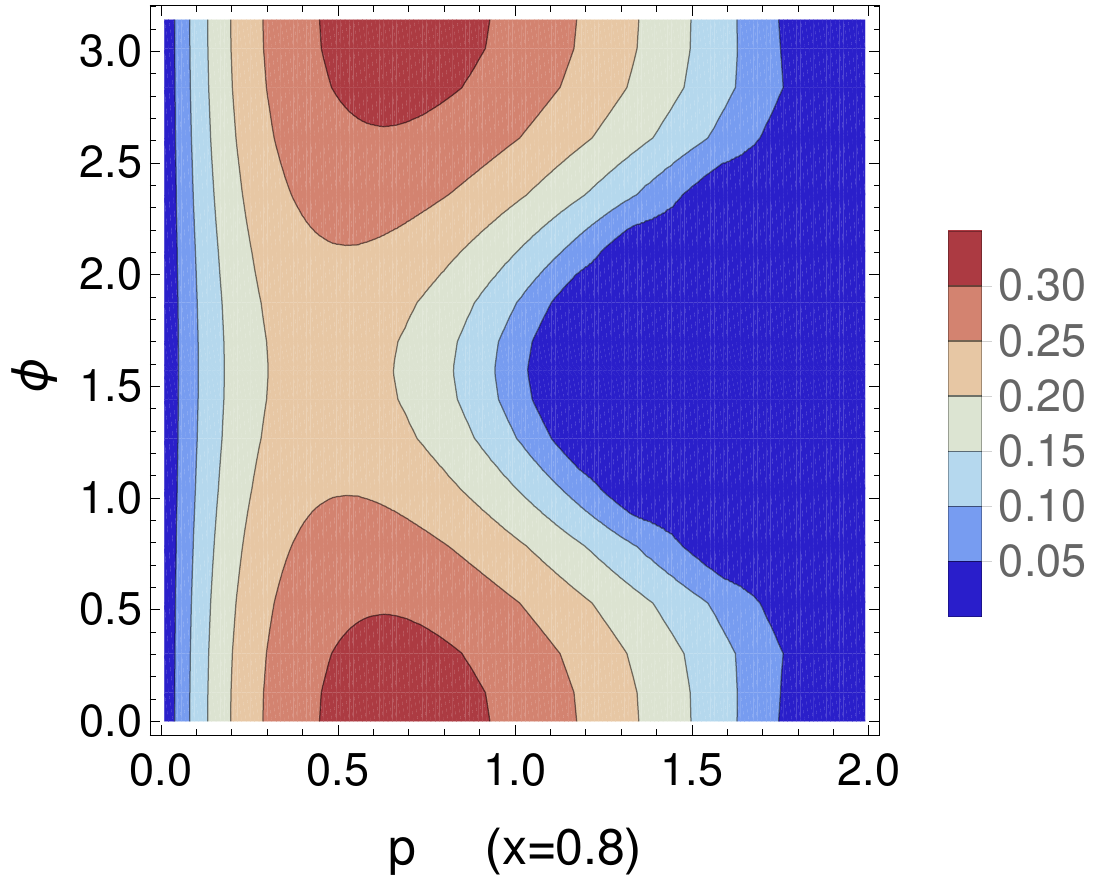} \quad
\caption{Contour plots of the largest imaginary solution from the distributions 6 (left two panels) and 7 (right two panels) with $\hat\mu_5=0$. \label{fig-meg8-aa}}
\end{figure}

\begin{figure}[htb]
\centering
\includegraphics[scale=.275]{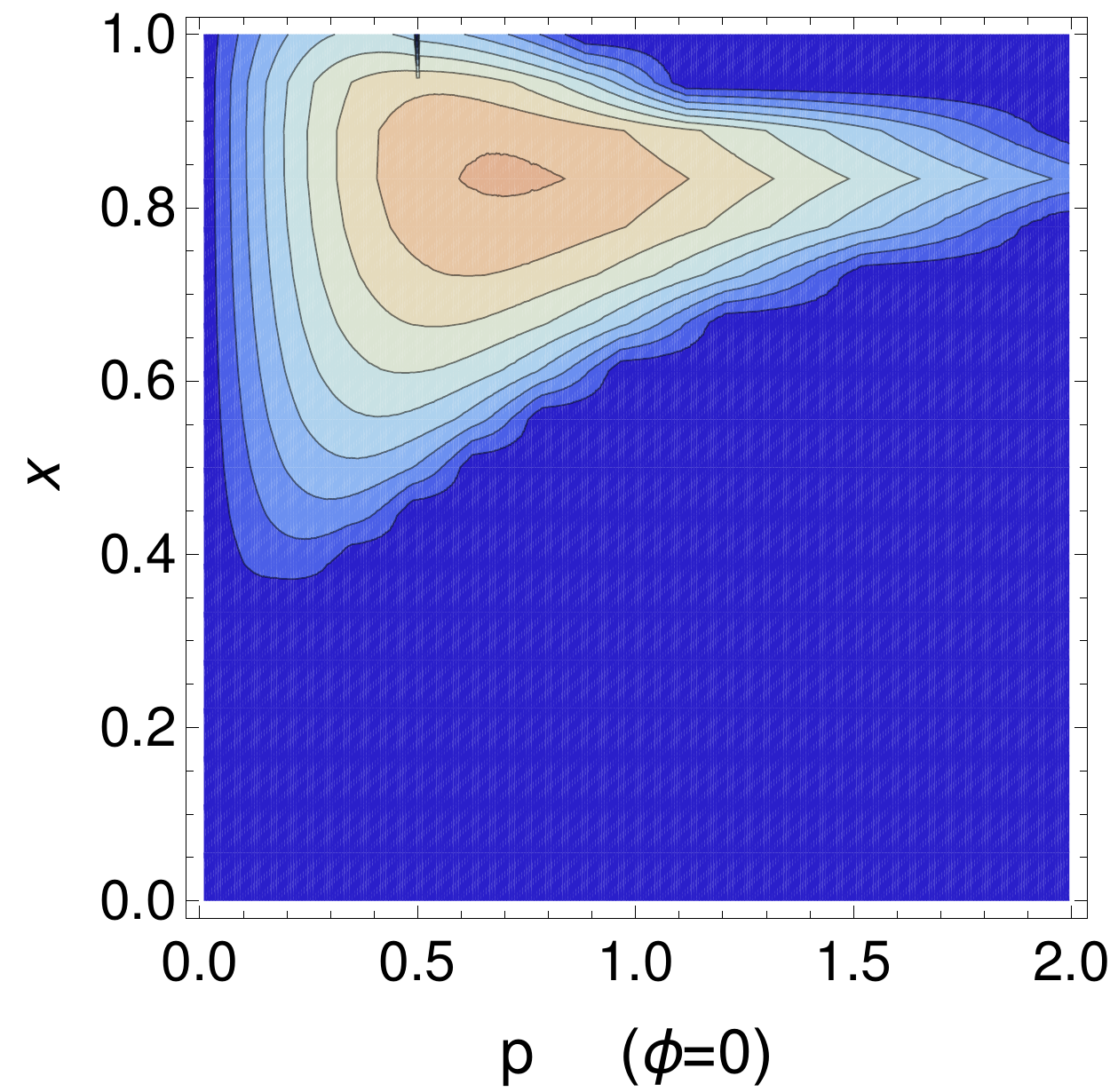} \quad
\includegraphics[scale=.275]{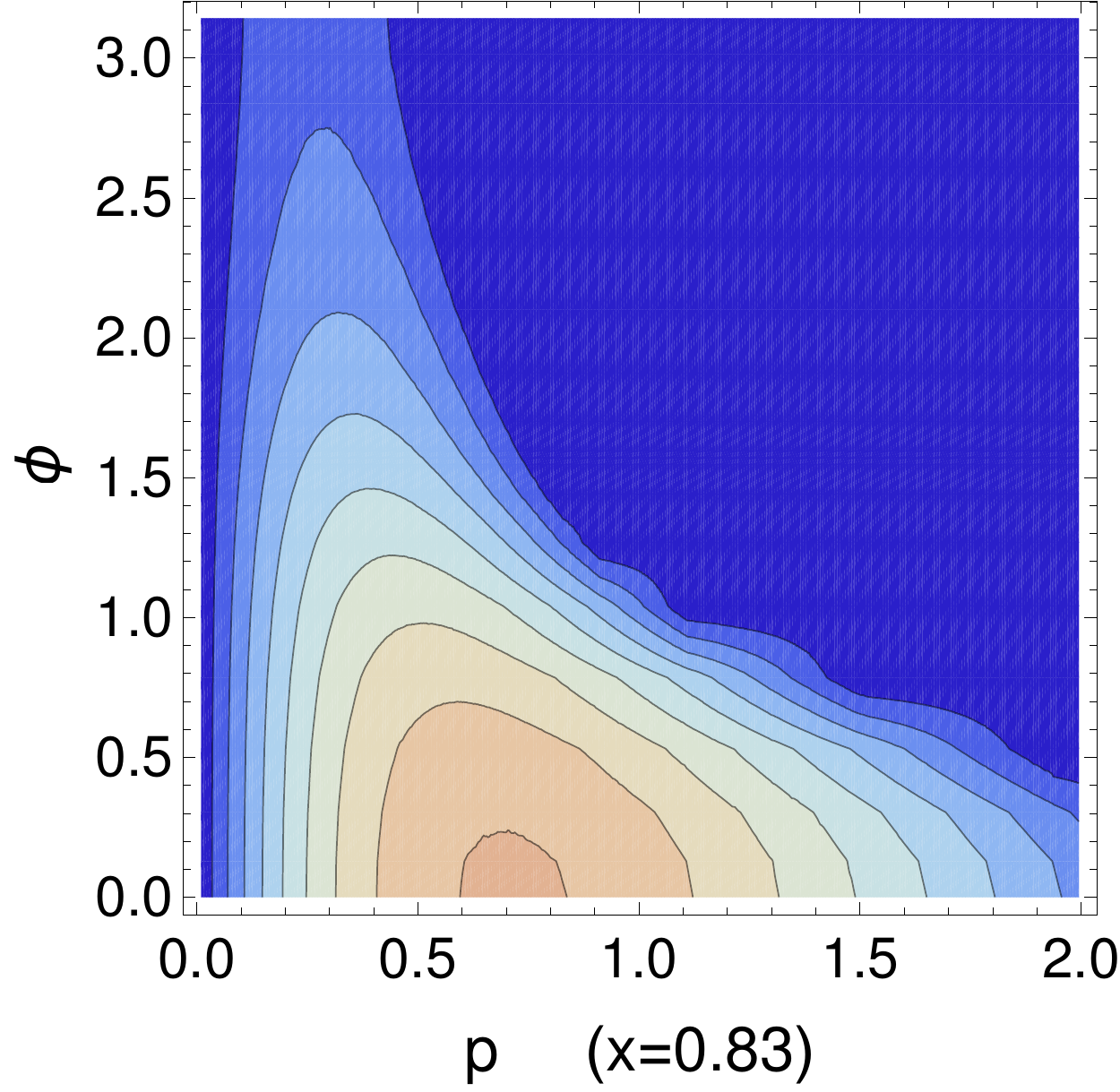} \quad
\includegraphics[scale=.275]{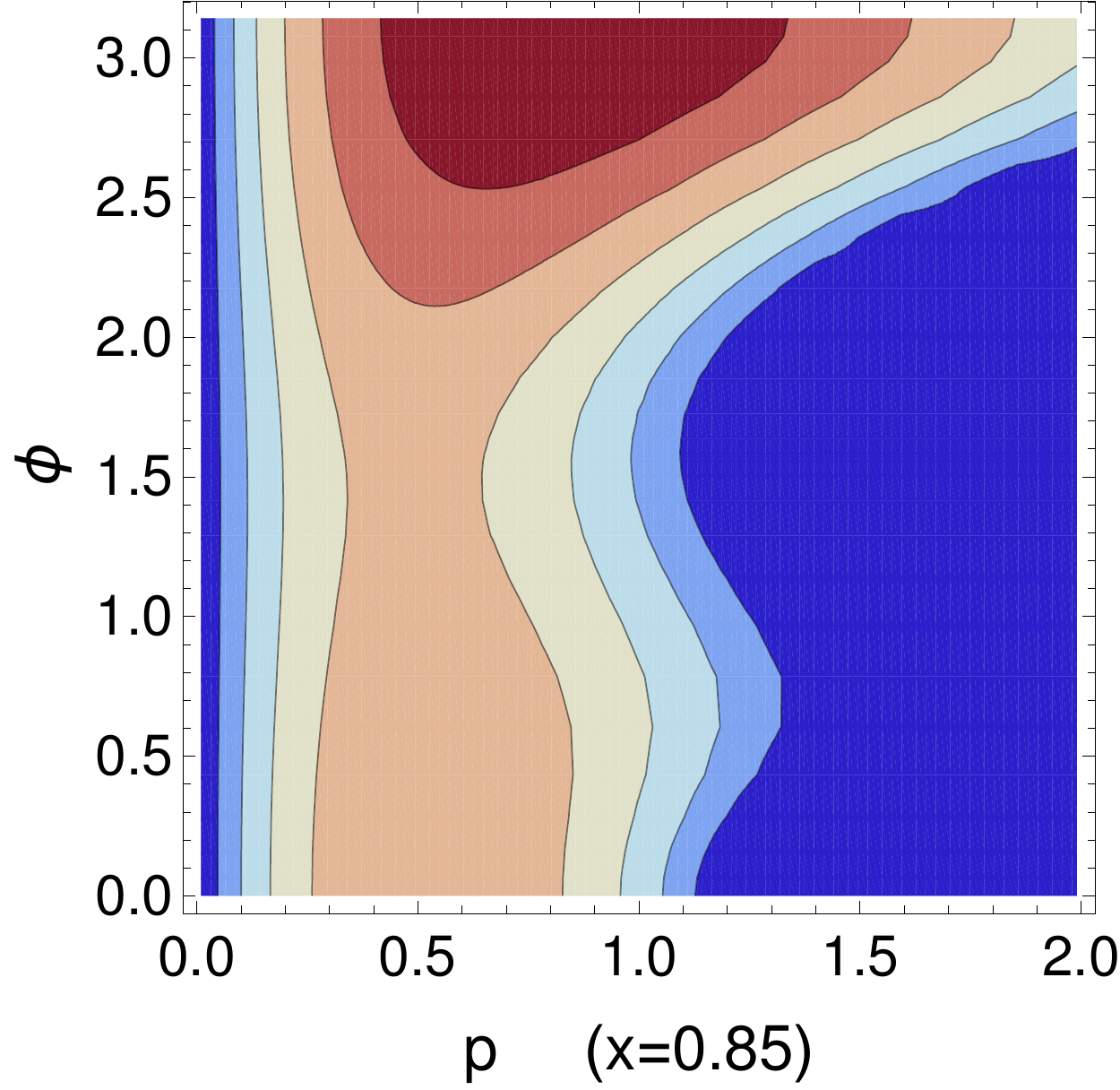} \quad
\includegraphics[scale=.38]{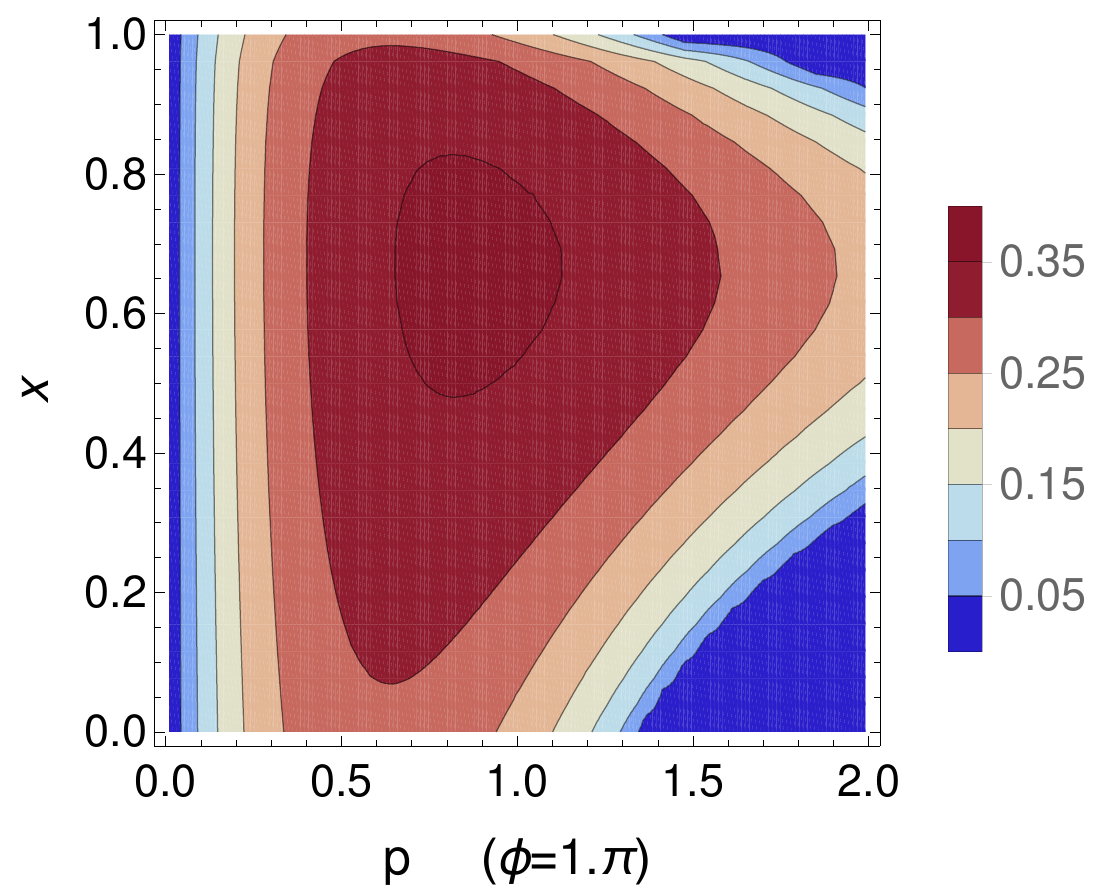} \quad
\caption{Contour plots of the largest imaginary solution from the distributions 9 (left two panels) and 8 (right two panels) with $\hat\mu_5=0$. \label{fig-meg4-aa}}
\end{figure}

The distributions 6-9 involve anisotropy parameters that correspond to higher spherical harmonics, and have a much richer structure. 
In figures \ref{fig-meg8-aa} and \ref{fig-meg4-aa} we show some contour plots of the largest imaginary solution obtained from these distributions with $\hat\mu_5=0$, for different choices of $x=\cos(\theta)$ and $\phi$.

The effect of the anisotropy parameters which correspond to terms with $e_1$ and $e_3$ individually odd (see the discussion at the beginning of section \ref{sec-YI}) can be seen by calculating $s_2\equiv \int_0^\infty dp \int^1_0 dx \int^\pi_0 d\phi \,  \sin(2\phi) \gamma(p,x,\phi)$. 
Any distribution for which $\xi_6=\xi_8=\xi_{13}=0$ will give $s_2=0$, which corresponds physically to symmetry across the reaction plane. 
If any of these anisotropy parameters are non-zero, the value of $s_2$ will also be non-zero. 
Physically this effect could be produced by a collision at non-zero impact parameter of two ions of different size. 
Distribution 6 and 7, for which we should find $s_2=0$, give $s_2 \approx 9\times 10^{-10}$ and $s_2 \approx 7\times 10^{-8}$, which can be taken as a measure of the numerical error in our computation. 
Distributions 8 and 9 give, respectively,  $s_2=-0.059$ and $s_2=0.068$.

\begin{figure}[H]
\centering
\includegraphics[scale=.275]{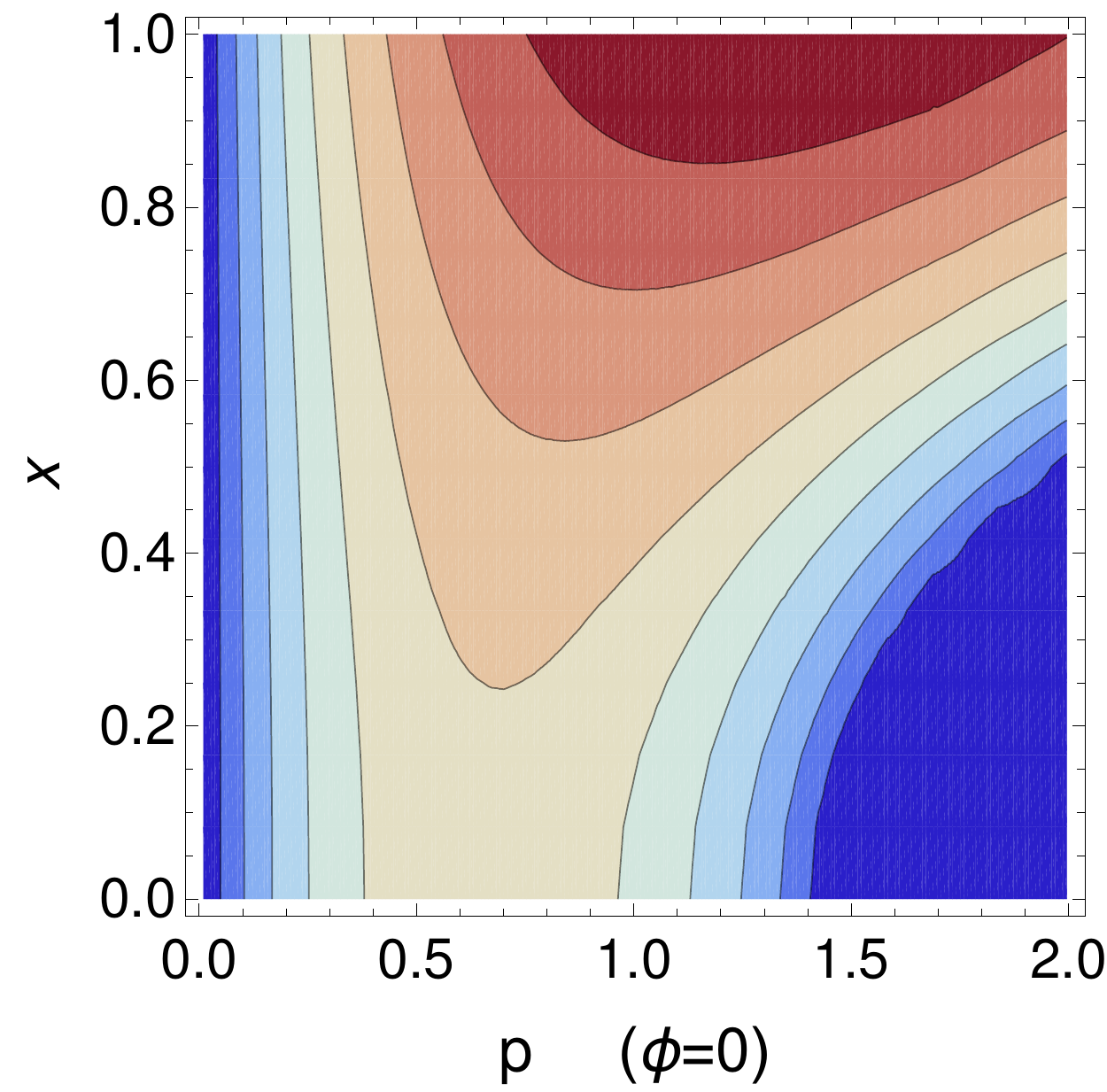} \quad
\includegraphics[scale=.275]{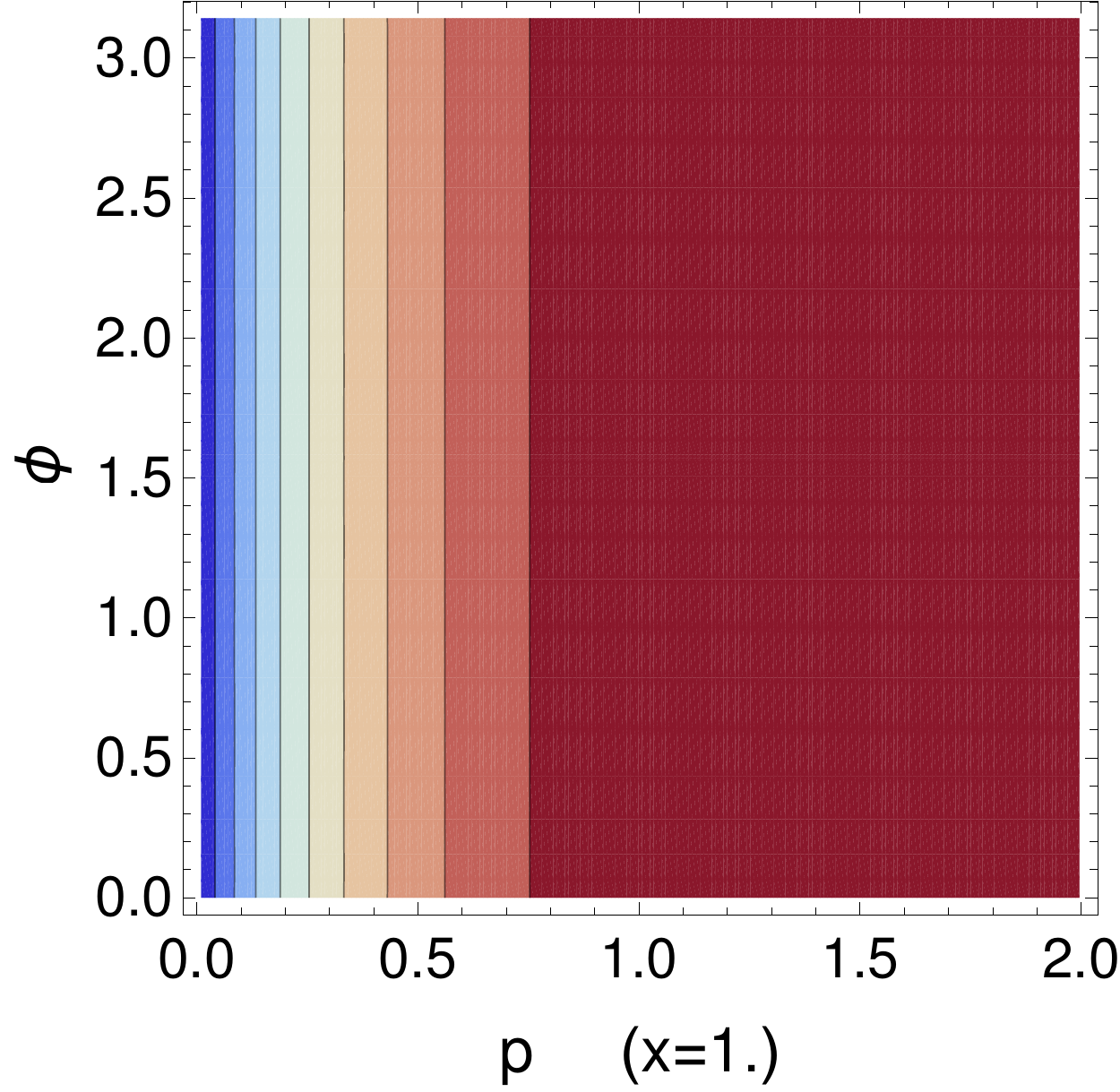} \quad
\includegraphics[scale=.275]{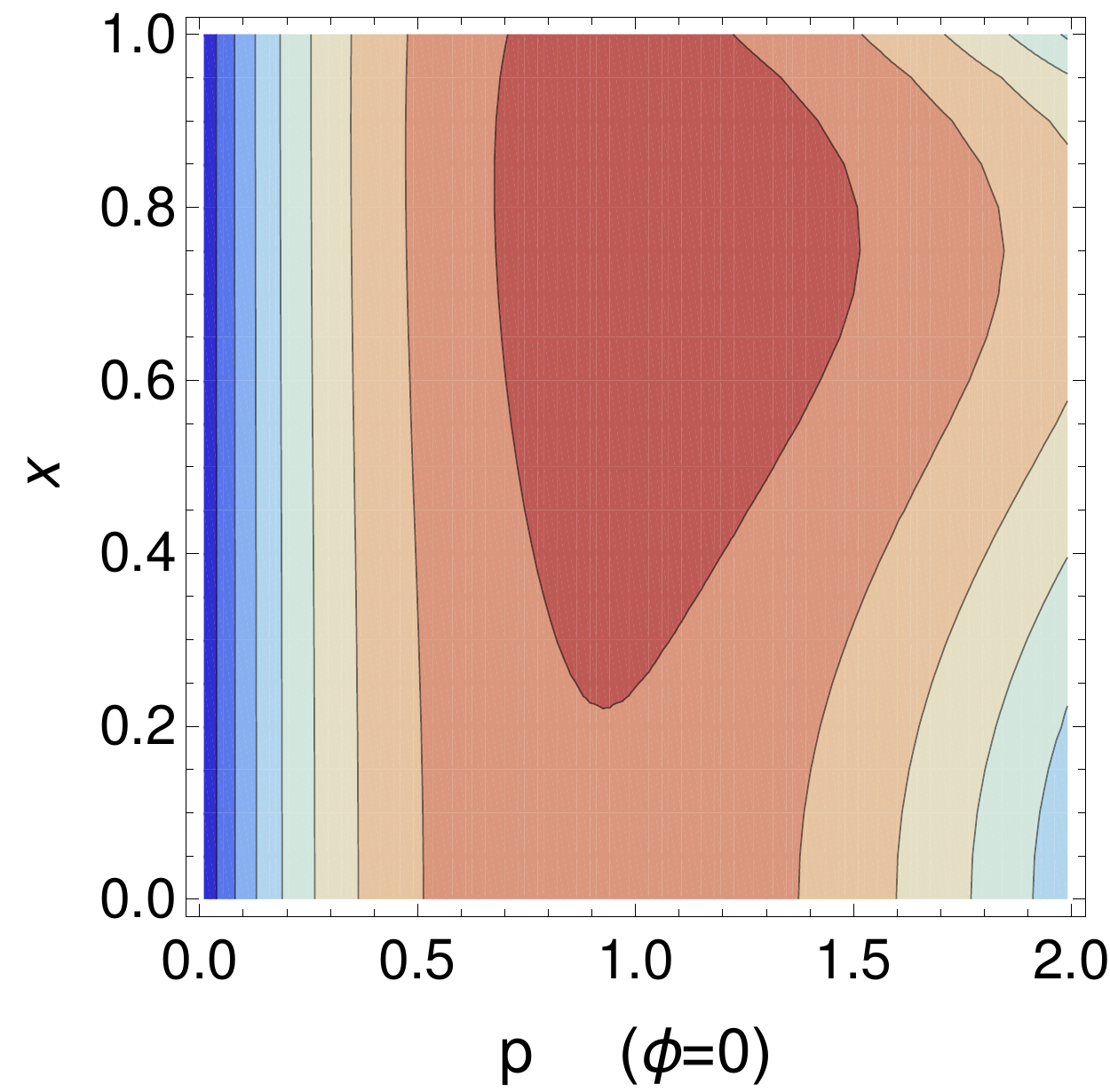} \quad
\includegraphics[scale=.38]{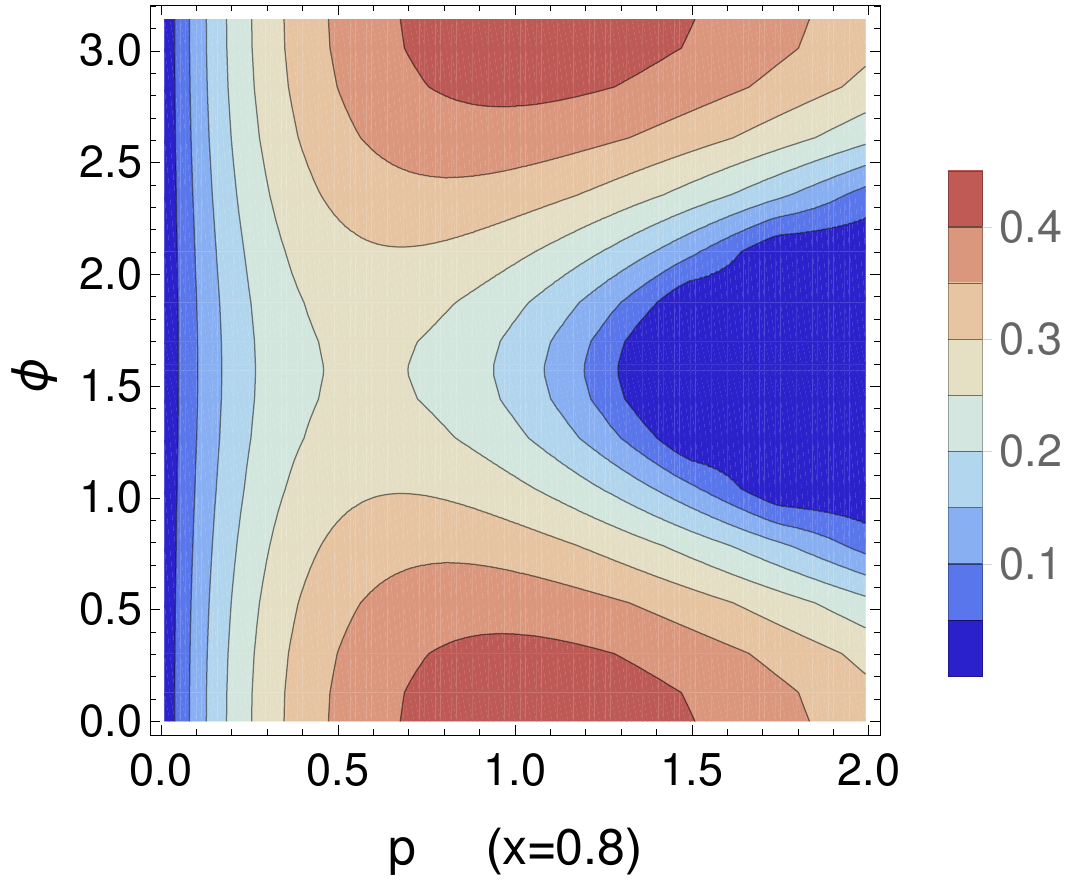} \quad
\caption{Contour plots of the largest imaginary solution for the distributions 6 (left two panels) and 7 (right two panels) with $\hat\mu_5=0.3$. \label{fig-mu5}}
\end{figure}

Finally, to show the effect of $\hat\mu_5\ne 0$, we plot in figure \ref{fig-mu5} the solutions obtained with $\hat\mu_5$=0.3 using the distributions 6 and 7. Comparison with figure \ref{fig-meg8-aa} shows that the chiral chemical potential increases the size and domain of the imaginary solution, but not uniformly.  

\section{Conclusions}

We have studied plasmons in anisotropic plasmas using a hard loop approach within the quasi-particle regime. 
We have used more general distribution functions than have been previously considered, with a technique that could be easily generalized 
to account for the anisotropies that are relevant in any given physical situation. 
The distribution function we have used depends on 9 anisotropy parameters and is normalized so that at zero chemical potential the Debye mass parameter is the same for all distributions. 
In an ultra-relativistic plasma at zero chemical potential this parameter is the only dimensionful scale. It is therefore meaningful to compare the solutions obtained from the dispersion equations for different choices of anisotropy parameters. 

Imaginary solutions to the dispersion equation are important because they are associated with plasma instabilities, which have the potential to strongly influence plasma dynamics. 
The important effect of imaginary collective modes on physical quantities in anisotropic plasmas, like transport coefficients, production rates, and bound state properties, is well documented in the literature. 
We have shown that the more general distribution functions we have introduced can significantly increase imaginary solutions of the dispersion equation in terms of both the magnitude of the mode, and the domain of wave vectors over which it exists. 
The distribution that is most commonly used in the literature depends on only one anisotropy parameter, which describes the oblateness of a distribution that is squeezed in one direction. 
An example is our distribution 3. 
In comparison,  our distribution 7, which has approximately the same value of the parameter $\xi$ that characterises its oblateness (see table \ref{table-choices}), produces a largest imaginary mode that is almost twice as big, and the integral of the solution over the momentum phase space is approximately 10 times larger. 
We have also shown that the effect of the chiral chemical potential is greatly enhanced by anisotropy. This is seen clearly from the results in table \ref{table-choices} which show that, relative to the isotropic result, the imaginary modes are much greater when 
even very moderate anisotropy is introduced. 

\acknowledgments
This work was supported by the Natural Sciences and Engineering Research Council of Canada Discovery  Grant  program. 

\appendix

\section{Retarded polarization tensor}
\label{retarded-all}

In this Appendix we calculate the one-loop retarded polarization tensor at finite temperature and chemical potential with the real-time formulation of finite temperature field theory, using the Keldysh representation \cite{Kobes:1984vb,LeBellac:1997rw,Carrington:2006xj}.

The electron propagator is a 2$\times$2 matrix of the form
\bea
G & = & \left(\begin{array}{cc}
G_{rr} &G_{ra} \\
G_{ar} & G_{aa} \\
\end{array}\right)
= \left(\begin{array}{cc}
G_{\rm sym} &G_{\rm ret} \\
G_{\rm adv} & 0 \\
\end{array}\right) 
\eea
where the retarded, advanced and symmetric propagators are given by
\bea
\nn
&& G_{\rm ret}(p_0,p)  = (\slashed{P} + m)r(p_0,p)
\\ 
\label{prop-def1}
&& G_{\rm adv}(p_0,p) = (\slashed{P} + m) a(p_0,p)
\\ \nn
&& G_{\rm sym}(p_0,p) = (\slashed{P} + m) f(p_0,p)
\eea
with
\bea
\nn
r(p_0,p)  &\equiv& \frac{1}{P^2-m^2+i0^+ {\rm sgn}(p_0)} 
\\ [2mm]
\label{prop-def2}
a(p_0,p) &\equiv& \frac{1}{P^2-m^2-i0^+ {\rm sgn}(p_0)} 
\\ [2mm] \nn
 f(p_0,p) &\equiv& -2\pi i 
\big[ \big(1-2 n^+(p_0) \big) \Theta(p_0) +  \big(1-2 n^-(p_0) \big) \Theta(-p_0) \big]
\delta(P^2 - m^2) \label{f-sym}
\eea
where 
\bea
n^+(p_0) = \frac{1}{e^{\beta(p_0-\mu)}+1} \text{~~ and ~~}   n^-(p_0) = \frac{1}{e^{-\beta(p_0-\mu)}+1}\,.
\label{distro-def}
\eea  
The self-energy has the form
\bea
\Pi &=& \left(\begin{array}{cc}
\Pi_{rr} & \Pi_{ra} \\
\Pi_{ar} & \Pi_{aa} 
\end{array}\right)
= \left(\begin{array}{cc}
0 & \Pi_{\rm adv} \\
\Pi_{\rm ret} & \Pi_{\rm sym} 
\end{array}\right)
\eea
and the vertex function is a 2$\times$2$\times$2 tensor which can be written 
\bea
\Gamma^\mu &=& -ie \gamma^\mu \left(
\begin{array}{cc}
 \{\Gamma_{rrr},\Gamma_{rra}\} & \{\Gamma_{rar},\Gamma_{raa}\} \\
 \{\Gamma_{arr},\Gamma_{ara}\} & \{\Gamma_{aar},\Gamma_{aaa}\} \\
\end{array}
\right) = -ie \gamma^\mu \left(
\begin{array}{cc}
 \{0,1\} & \{1,0\} \\
 \{1,0\} & \{0,1\} \\
\end{array}
\right) .
\eea

The contribution to the retarded self-energy from the one-loop diagram is
\bea
i \Pi_{\rm ret}^{\mu \nu} (p_0,p) = i \Pi_{ar}^{\mu \nu}(p_0,p) = \frac{(-ie)^2}{2}\sum_{ii'jj'}\int \frac{d^4k}{(2\pi)^4} \,
\gamma^\mu \Gamma_{aij} \, G_{jj'}(k) G_{i'i}(k-p) \, \gamma^\nu \Gamma_{ri'j'} . 
\eea
The sum over Keldysh indices $\{i,i',j,j'\}\in\{r,a\}$ is easily done because $G_{aa}=0$ and a vertex function with an odd number of $a$ indices vanishes.  The result is 
\bea
\label{A5} 
\Pi_{\rm ret}^{\mu\nu} (p_0,p) 
= i\frac{e^2}{2}
\int \frac{d^4k}{(2\pi)^4} \,
{\rm Tr}\big[\gamma^\mu (\slashed{K} + m) \gamma^\nu (\slashed{K}-\slashed{P} + m)\big] 
\big[r(K) f(K-P) + f(K) a(K-P)\big] \,\nn\\\label{pi-ret}
\eea
where we have used the short hand notation $r(K)$ to represent $r(k_0,k)$, and similarily for other propagator components. From this point on we will suppress the subscript that indicates the retarded component of the polarization tensor. In addition we consider only the relativistic limit $T\gg m$ and therefore we drop the electron mass in the integrand in equation (\ref{A5}). We also introduce the short-hand notation $dK\equiv d^4k/(2\pi)^4$. 

We will consider a chirally asymmetric plasma which is characterised by a difference in the chemical potentials of the right and left handed fermions, which are denoted $\mu_R$ and $\mu_L$. We define
\bea
\label{A5a} 
&& \Pi_R^{\mu\nu} (p_0,p) 
= \frac{ie^2}{2}
\int dK \,
{\rm Tr}\big[(1+\gamma^5)\gamma^\mu \slashed{K}  \gamma^\nu (\slashed{K}-\slashed{P})\big] 
\big[r(K) f(K-P) + f(K) a(K-P)\big]\bigg|_{\mu=\mu_R}  \nn\\
&& \Pi_L^{\mu\nu} (p_0,p) 
= \frac{ie^2}{2}
\int dK \,
{\rm Tr}\big[(1-\gamma^5)\gamma^\mu \slashed{K} \gamma^\nu (\slashed{K}-\slashed{P})\big] 
\big[r(K) f(K-P) + f(K) a(K-P)\big] \bigg|_{\mu=\mu_L}\, \nn \\ \label{pi-rl}
\eea
and 
\bea
&& \Pi^{\mu\nu}(p_0,p) = \frac{1}{2}\left(\Pi_R^{\mu\nu} (p_0,p) +\Pi_L^{\mu\nu} (p_0,p)\right)  \,.\label{pi-sum}
\eea
The integrals in equations (\ref{pi-rl}) can be rewritten by performing a shift of the 4-vector $K \to K+P$ on the term in square brackets that does not have a factor $f(K)$. All terms in the resulting expression have a factor $f(K)$ and, using equation (\ref{f-sym}), this factor is proportional to $\delta(K^2)$ in the relativistic limit. The integral over $k_0$ can be done using this delta function. 
We work in time-like axial gauge and calculate only the spatial components of the polarization tensor. After doing the $k_0$ integral we obtain the results in equations (\ref{pi-even}, \ref{pi-odd}).

\section{Nyquist analysis}
\label{nyquist-sec}

The dispersion equation that we have solved numerically is ${\cal D}=0$ where ${\cal D}$ is given in equation (\ref{den9}). 
When numerical methods are used to find the solutions of an equation, one must input the range over which the search will be performed, and solutions outside this range will be missed. 
A Nyquist analysis can be used to determine the number of solutions that exist so that the search range can be extended as needed. 

To explain the idea we discuss a generic equation of the form
\bea
\label{general-eq}
f(p_0) = 0
\eea
and  define the function 
\bea
\label{F-def}
F(p_0) \equiv \frac{f^\prime(p_0)}{f(p_0)} = \frac{d}{dp_0}{\rm ln}f(p_0) \,.
\eea
We consider the contour integral 
\bea
\label{Nyq-int-1}
\oint_C \frac{dp_0}{2\pi i} F(p_0) 
\eea
where the contour is a positively (counter-clockwise) oriented closed loop, which is chosen so that $F(p_0)$ is analytic inside the loop except at isolated points. The integral is equal to the sum of the residues. The residue of $F(p_0)$ at a zero of $f(p_0)$ of order $l$ is $l$, and the residue of $F(p_0)$ at a pole of $f(p_0)$ of order $l$ is $-l$. This gives 
\bea
\label{Nyq-int-2}
\oint_C \frac{dp_0}{2\pi i} F(p_0) = n_Z - n_P 
\eea
where $n_Z$ and $n_P$ are the numbers of zeros and poles of $f(p_0)$ inside the contour $C$, taking into account the fact that each zero and pole of order $l$ is counted $l$ times. The purpose of the Nyquist analysis is to determine $n_Z$.

\begin{figure}[htb]
\center
\includegraphics*[width=0.55\textwidth]{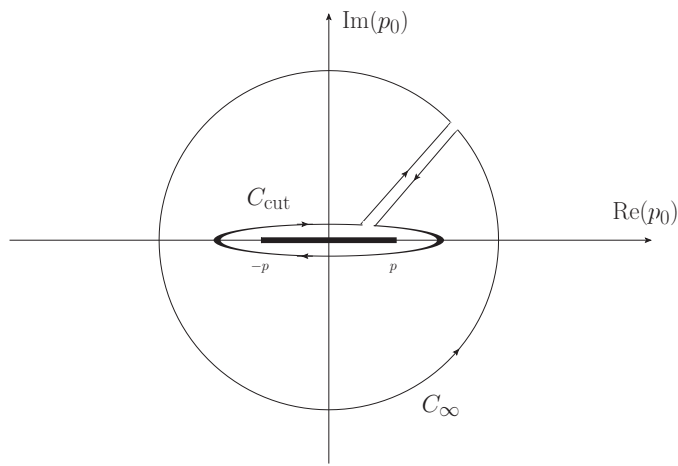}
\caption{The contour in the plane of complex $\hat p_0$ which is used to compute the number of solutions of the dispersion equations.} 
\label{fig-Nyquist-1}
\end{figure}

The anisotropic dressing functions are obtained numerically from the integrals in equations (\ref{pi-even-3}, \ref{pi-odd-3}). 
 Using $P\cdot V = p(\hat p_0 - x')$ (see equations (\ref{new-coords}, \ref{new-v})) one finds that the dressing functions have a cut along  the real axis for $|\hat p_0|<1$, and the dispersion equation therefore has a cut for $|p_0|<p$. 
This means that the contour $C$ can be chosen as in Fig.~\ref{fig-Nyquist-1}. 
The integrals along the lines connecting the circular contour $C_\infty$ to $C_{\rm cut}$ always compensate each other and therefore the contour integral (\ref{Nyq-int-2}) equals
\bea
\label{Nyq-int-LM}
\oint_{C_\infty} \frac{dp_0}{2\pi i} F(p_0) +
\oint_{C_{\rm cut}} \frac{dp_0}{2\pi i} F(p_0) = n_Z - n_P\,.
\eea
The contribution from the big circle is calculated by writing $p_0 = |p_0|e^{i \phi}$ and taking $|p_0|\to\infty$ which gives  
\bea
\label{Nyq-int-infty}
\oint_{C_\infty} \frac{dp_0}{2\pi i} F(p_0) = \lim_{|p_0| \rightarrow \infty} p_0 F(p_0) \equiv n_\infty \,.
\eea
The integral along the cut can be calculated using that $F(p_0)$ is the logarithmic derivative of $f(p_0)$ which gives 
\bea
\label{Nyq-int-cut}
\oint_{C_{\rm cut}} \frac{dp_0}{2\pi i} F(p_0)  
= \frac{1}{2\pi i}\oint_{C_{\rm cut}}  \frac{d}{dp_0}{\rm ln}f(p_0)
= \frac{1}{2\pi i} \Big( {\rm ln} f(p_{0{\rm end}}) - {\rm ln} f(p_{0{\rm start}}) \Big) \equiv n_W 
\eea
where $p_{0{\rm start}}$ is the (arbitrarily chosen) starting point of the contour which encloses the cut, and $p_{0{\rm end}}$ is the end point. 
The quantity $n_W$ is the number of times that the curve in the plane of complex $f$ 
travels counter-clockwise around the point $f=0$, and is called a winding number. 
Combining the results (\ref{Nyq-int-infty}, \ref{Nyq-int-cut}), we have that equation (\ref{Nyq-int-LM}) gives
\bea
\label{ny-all}
n_Z = n_P + n_\infty + n_W \,.
\eea

A Nyquist analysis of the equation ${\cal D}=0$ can be done analytically in the weakly anisotropic limit. 
When $\mu_5=0$ the dispersion equation in the weakly anisotropic limit factors into three separate equations
\bea
&& f^{(1)}(p_0,\vec p) =  P^2-\pi_1(p_0,\vec p)=0 \label{anio-pi1}\\
&& f^{(2)}(p_0,\vec p) = \frac{1}{p_0^2}\left(p_0^2-\pi_2(p_0,\vec p)\right) = 1-\frac{\pi_2(p_0,\vec p)}{p_0^2} =0 \label{anio-pi2}\\
&& f^{(3)}(p_0,\vec p) = P^2-\pi_3(p_0,\vec p)=0 \label{anio-pi3}\,.
\eea
When $\mu_5\ne 0$ equations (\ref{anio-pi1}, \ref{anio-pi3}) are replaced by the weakly anisotropic versions of equations (\ref{Pmodes}, \ref{Mmodes}) which are
\bea
&& f^{(4)}(p_0,\vec p) = P^2-\left(\frac{1}{2}(\pi_1(p_0,\vec p)+\pi_3(p_0,\vec p)) + \pi_8(p_0,\vec p)\right)=0 \label{anio-plus}\\
&& f^{(5)}(p_0,\vec p) = P^2-\left(\frac{1}{2}(\pi_1(p_0,\vec p)+\pi_3(p_0,\vec p)) -\pi_8(p_0,\vec p)\right)=0 \,.\label{anio-minus}
\eea
In all cases it is easy to see that $n_P=0$. 
In equation (\ref{anio-pi2}) we have divided out a zero mode, so that $n^{(2)}_\infty = 0$, and for all of the other dispersion equations it is straightforward to see that $n_\infty = 2$.

\begin{figure}[htb]
\center
\includegraphics*[width=0.51\textwidth]{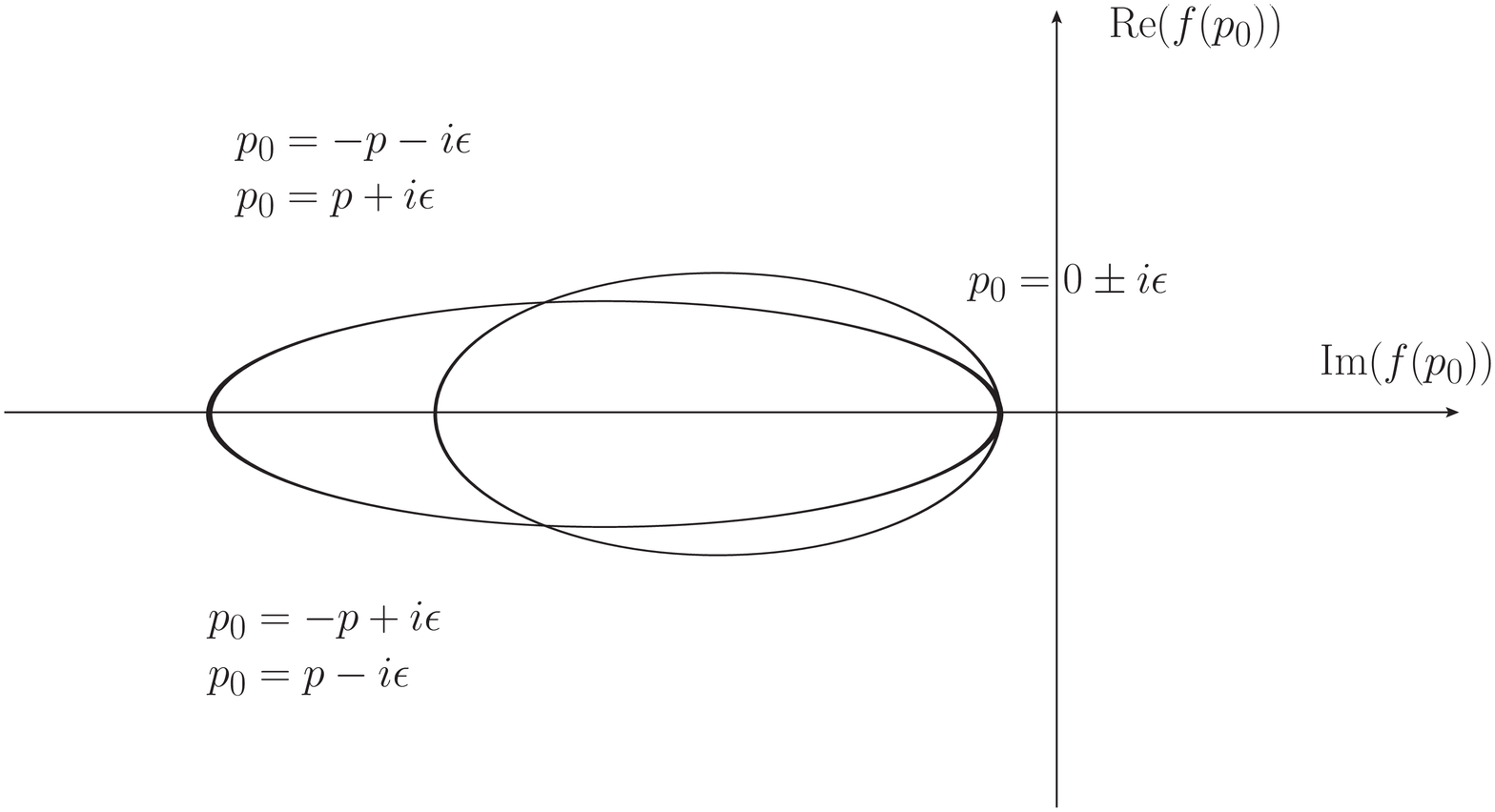}
\includegraphics*[width=0.48\textwidth]{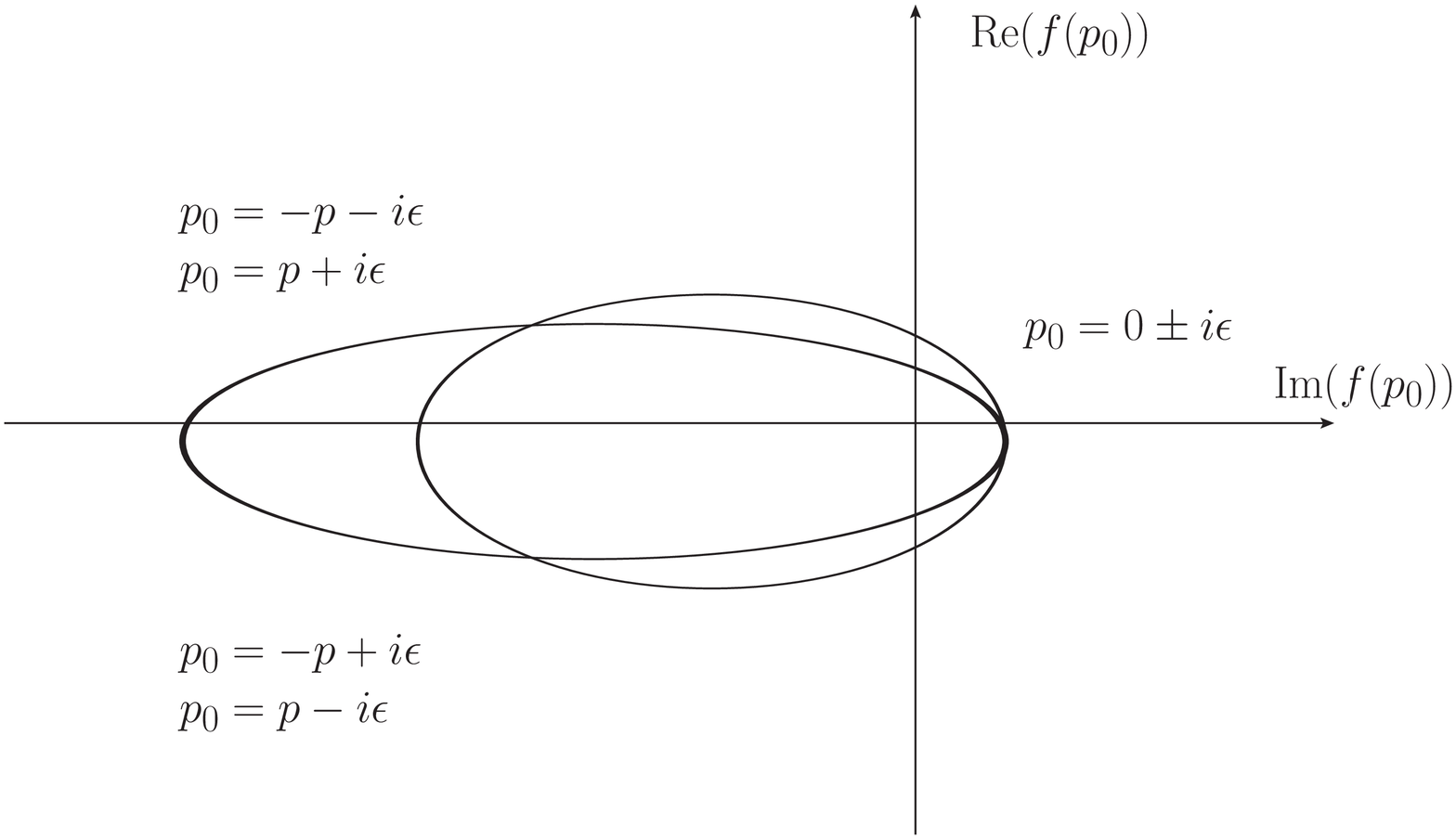}
\caption{Mappings of the contour $C$ in the plane of complex $p_0$. In the right panel the corresponding winding number is two, and in the left panel it is zero. } 
\label{nyquist-iso}
\end{figure}

The non-trivial part of the calculation is the determination of the winding number, which is found by mapping the closed contour $C_{\rm cut}$ in the plane of complex $p_0$ onto a path in the plane of complex $f(p_0)$. 
The points $p_{0{\rm start}}$ and $p_{0{\rm end}}$ in equation (\ref{Nyq-int-cut}) have the same modulus, but their phases differ by $2\pi$. 
The winding number equals the number of times that the curve in the plane of complex $f$ 
travels counter-clockwise around the point $f=0$ (some detailed examples of these mappings can be found in Ref. \cite{Carrington:2014bla}).
There are two possibilities, which are shown in figure \ref{nyquist-iso}. 
In the right panel, the mapping circles the origin twice, and the winding number is 2, while in the left panel the winding number is zero.
To distinguish between the two possibilities for each dispersion equation, we need to determine the sign of Re[$f(0,p)$].

We discuss first the dispersion equation (\ref{anio-pi2}), which is different from the other four. 
In the weakly anisotropic limit we have
\bea
f^{(2)}(0,\vec p) = 2+\tilde f(\xi_i,x,\phi)
\eea
where $\tilde f(\xi_i,x,\phi)$ is a dimensionless function of the angular variables that depends linearly on the anisotropy parameters. 
Since we have assumed weak anisotropy we have $f^{(2)}(0,\vec p)$ always positive, which means $n_W=2$, and therefore equation (\ref{ny-all}) gives  $n_Z=2$.
These two solutions are always present, and numerically we find they are always real. They are anisotropic modifications of the HTL longitudinal modes (denoted $\pm\omega_L(\omega)$). 

For all of the remaining dispersion equations the number of solutions  is 
\bea
n_Z=2+n_W\,.
\eea
For equations (\ref{anio-pi1}) and (\ref{anio-pi3}) the winding number is determined from the sign of the functions
\bea
&& f^{(1)}(0,\vec p) = -\left(p^2- p^2_{1{\rm crit}}(\xi_i,x,\phi)\right) \label{more1}\\
&& f^{(3)}(0,\vec p) = -\left(p^2- p^2_{3{\rm crit}}(\xi_i,x,\phi)\right)\,. \label{more3}
\eea
When the chirality parameter $\mu_5$ is non-zero we consider equations (\ref{anio-plus}, \ref{anio-minus}) instead of equations (\ref{anio-pi1}, \ref{anio-pi3}). 
For weak anisotropy the winding number is determined from the sign of the functions
\bea
&& f^{(4)}(0,\vec p) = -\left(p +  p_{5{\rm crit}}(\xi_i,x,\phi)\right) \\
&& f^{(5)}(0,\vec p) = -\left(p - p_{5{\rm crit}}(\xi_i,x,\phi)\right)\,.
\eea
The results for $p^2_{1{\rm crit}}$, $p^2_{3{\rm crit}}$ and $p_{5{\rm hat}}$ are given in equations (\ref{kcrit1}, \ref{kcrit3}, \ref{kcrit8}). 
\bea
15 p_{1{\rm crit}}^2 &=& 
\xi _8 x \sqrt{1-x^2} \left(4 x^2-1\right) \cos ^3(\phi )
+\left(\xi _{11} \left(1-2 x^2\right)^2-5 \xi _2 \left(x^2-1\right)\right) \cos ^2(\phi ) \nonumber \\
&+& \frac{1}{2} x \sqrt{1-x^2} \cos (\phi ) \left(-9 \xi _8 \cos (2 \phi )+10 \xi _6+9 \xi _8+6 \xi
   _{13}-8 \xi _{13} x^2\right) \nonumber \\
&-& \sin ^2(\phi ) \left(5 \xi _2-3 \xi _4+\xi _{11}+9 \xi_4 \left(x^2-1\right) \cos (2 \phi )+9 \xi _4 x^2-3 \xi _{11} x^2\right) \nonumber \\
&+& x^2 \left(5 \xi _9+6 \xi _{14}-4 \xi _{14} x^2\right)+2 \xi _4 \left(-2 x^4+x^2+1\right) \cos^4(\phi ) \label{kcrit1}\\
15 p_{3{\rm crit}}^2 &=& -3 \xi _4 x^2 \sin ^2(2 \phi )+\xi _{11} \left(2 x^2-1\right) \sin ^2(\phi )
+4 \xi _8 x \sqrt{1-x^2} \left(4 x^2-1\right) \cos ^3(\phi ) \nonumber \\
&+& x \sqrt{1-x^2} \cos (\phi ) \left(-3 \xi _8 \cos (2 \phi ) + 10 \xi _6+3 \xi _8+4 \xi _{13} \left(3-4 x^2\right)\right)
+5 \xi _9 \left(2 x^2-1\right) \nonumber \\
&+& \cos
   ^2(\phi ) \left(6 \xi _4 \sin ^2(\phi )+5 \xi _2 \left(1-2 x^2\right)+2 \xi _{11}
   \left(8 x^4-8 x^2+1\right)\right) \nonumber \\
&+& 2 \xi _4 \left(-8 x^4+4 x^2+1\right) \cos ^4(\phi ) - 2 \xi _{14} \left(8 x^4-12 x^2+3\right)\label{kcrit3} \\
\frac{30}{\hat\mu_5} p_{5{\rm hat}} &=& 30+10 \xi _2+6 \xi _4-5 \xi _9-9 \xi _{14}+ 10 \sqrt{1-x^2} x \left(3 \left(\xi _6+\xi _8 x^2\right)
   +\xi _{13} \left(3-4 x^2\right)\right) \cos (\phi )\nonumber \\
   &-& 5 \left(x^2-1\right)
   \big(\xi _4 \left(x^2-1\right) \cos (4 \phi )
   -2 \xi _8 x \sqrt{1-x^2} \cos (3 \phi)   \nonumber \\ 
   && ~~~~~~~~~~ ~~~  +\left(3 \xi _2+2 \xi _4+\xi _{11}+4 \xi _4 x^2-4 \xi _{11} x^2\right) \cos (2 \phi)\big) \nonumber \\
&&   -5 x^2 \left(3 \left(\xi _2-2 \xi _9+\xi _4 x^2\right)
   +4 \xi _{14} \left(2x^2-3\right)\right)+\xi _{11} \left(20 x^4-15 x^2+2\right) \,. \label{kcrit8}
\eea   

In the isotropic case $p^2_{1{\rm crit}}(\xi_i,x,\phi) = p^2_{3{\rm crit}}(\xi_i,x,\phi) = 0$, the right side of equations (\ref{more1}, \ref{more3}) cannot be positive, and therefore the winding number is zero and both dispersion equations have two solutions. Remembering that in the isotropic limit $\pi_1=\pi_3=\Pi_T$, we see that these are the HTL transverse solutions ($\pm\omega_T(p)$). 
When non-zero anisotropy parameters are introduced, and when $p$ is small enough, there can be choices of the angles and anisotropy parameters for which $p^2_{1{\rm crit}}(\xi_i,x,\phi)>p^2>0$ and/or $p^2_{3{\rm crit}}(\xi_i,x,\phi)>p^2>0$. 
In this case, the winding number is two, which means that extra solutions to equations (\ref{anio-pi1}, \ref{anio-pi3}) exist. 

From equation (\ref{kcrit8}) we have that in the isotropic limit $p_{5{\rm crit}}(\xi_i,x,\phi)=\hat \mu_5$, which means that $n_W^{(4)}=0$ and $n_W^{(5)}=2$ if $p<\hat\mu_5$ and zero otherwise. Thus we have that for $\mu_5\ne 0$ an additional pair of solutions, which turn out to be imaginary, exist even in the isotropic limit (see figure \ref{disp-iso-fig}). The domain of these modes can be larger for certain orientations of the wave vector if the system is anisotropic. 

\begin{figure}[H]
\center
\includegraphics*[width=0.85\textwidth]{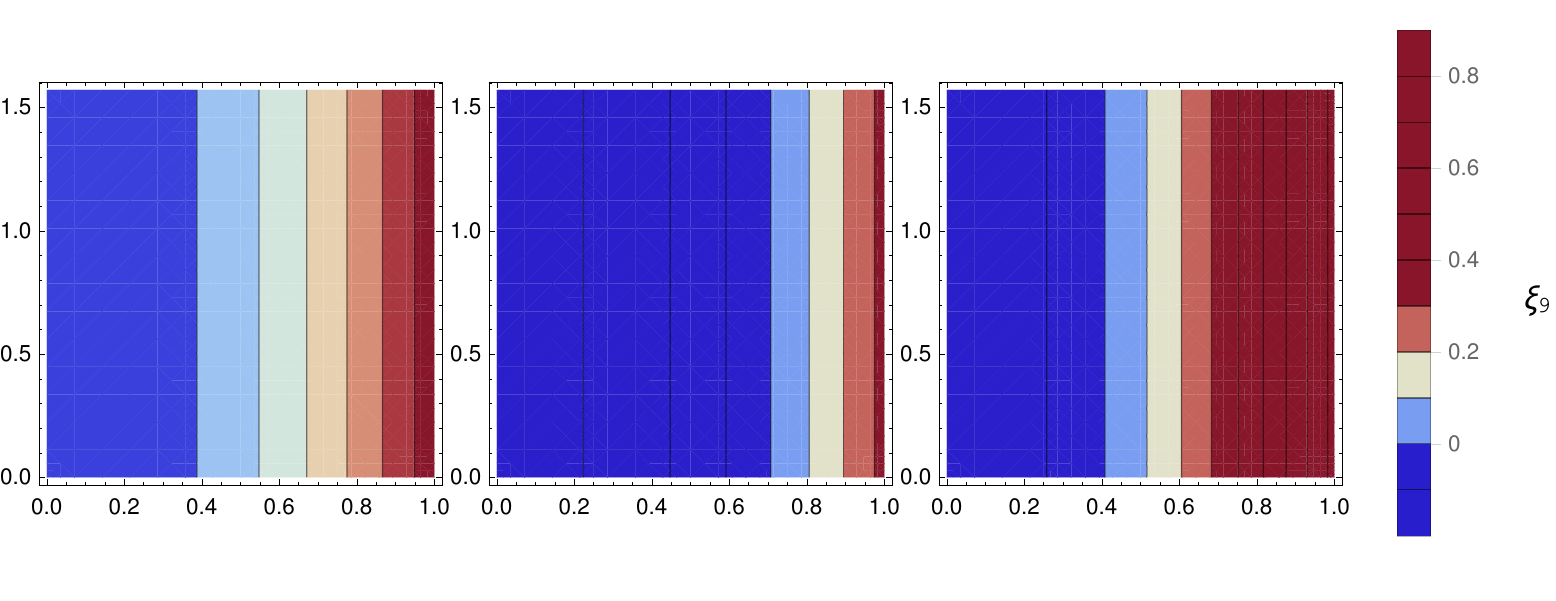}
\includegraphics*[width=0.85\textwidth]{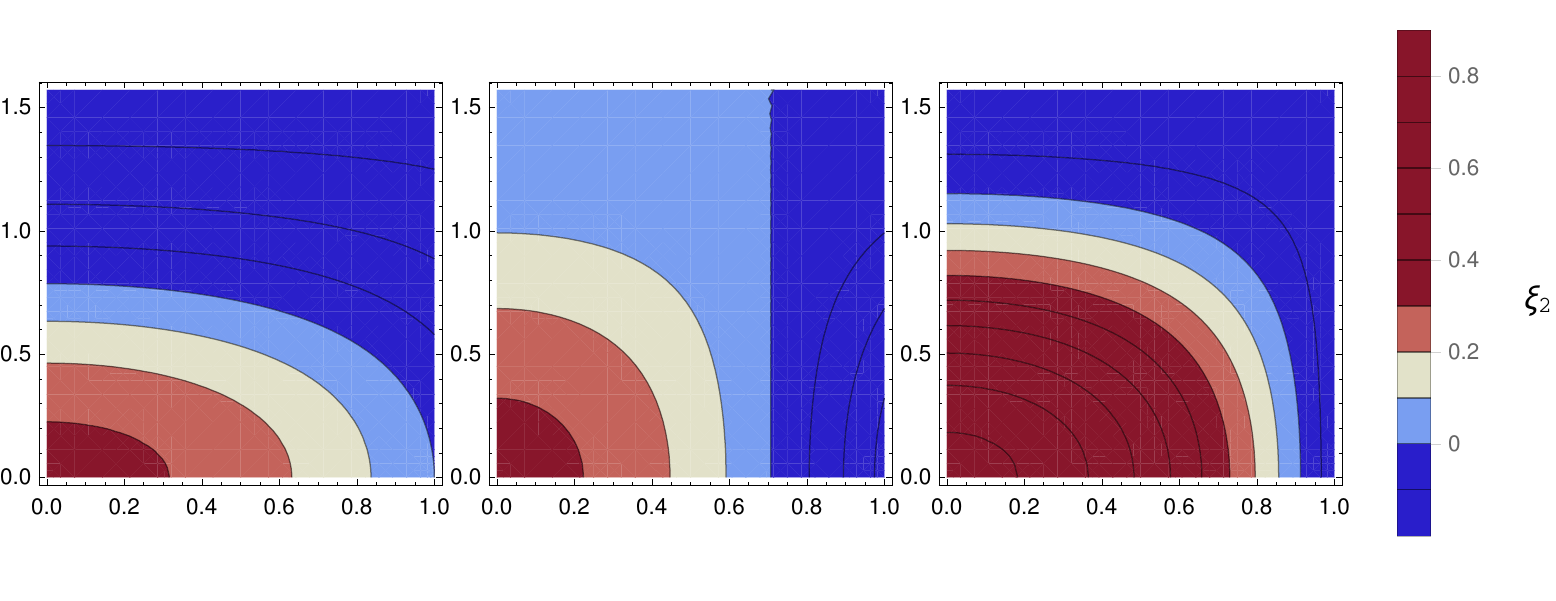}
\includegraphics*[width=0.85\textwidth]{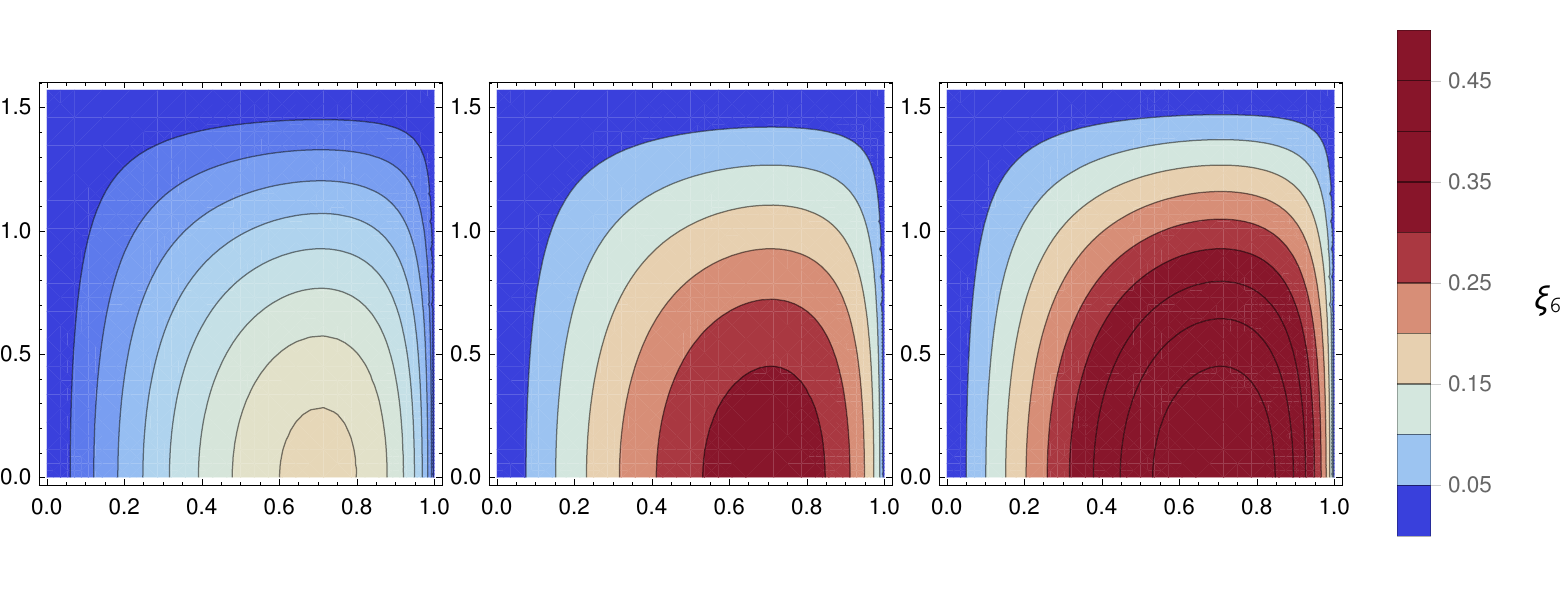}
\includegraphics*[width=0.85\textwidth]{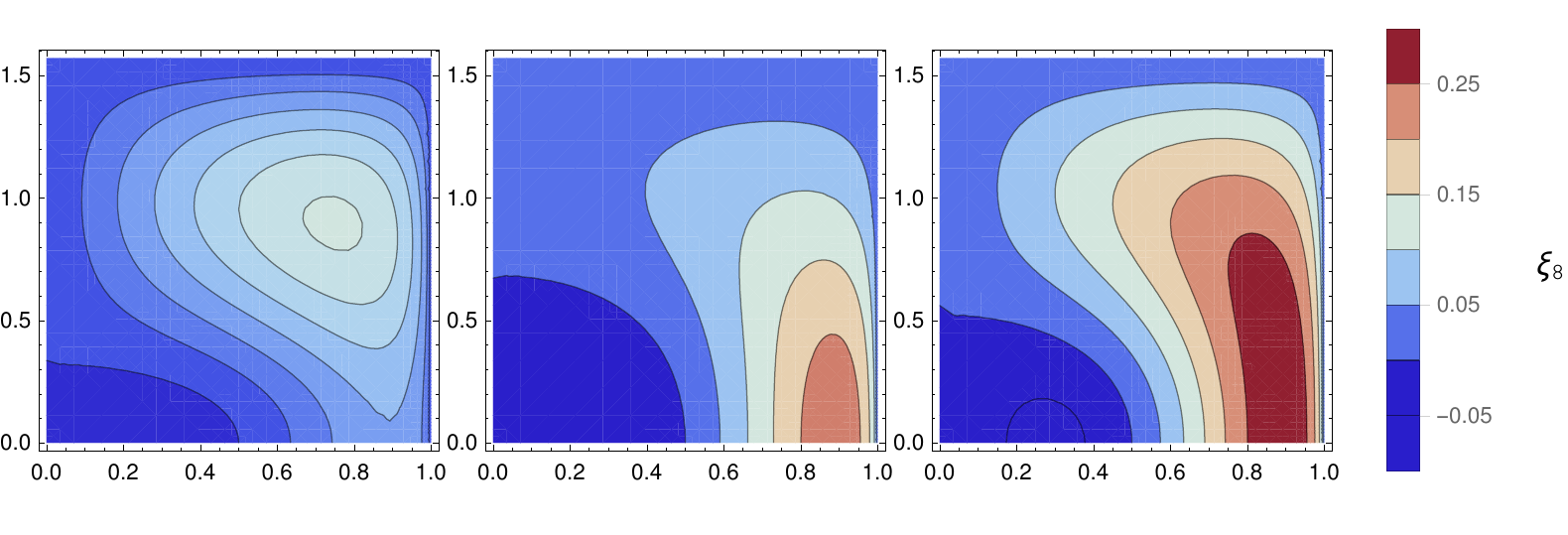}
\caption{Contour plots of the coefficients of the anisotropy parameters for $p_{1{\rm crit}}^2$ (left), $p_{3{\rm crit}}^2$ (center) and $p_{5{\rm crit}}$ (right).  The axes show $x=\cos(\theta)\in(0,1)$ and $\phi\in(0,\pi/2)$.   } 
\label{coefs-cont}
\end{figure}

The expressions in equations (\ref{kcrit1}, \ref{kcrit3}, \ref{kcrit8}) are only valid in the weakly anisotropic limit, however, we can use them to obtain information about the orientations of the wave vector $\vec p$ for which large imaginary solutions could be found with specific sets of anisotropy parameters. To obtain more insight, we can plot the coefficient of each anisotropy parameter in equations (\ref{kcrit1} - \ref{kcrit8}). We show the results for three of the coefficients in figure \ref{coefs-cont}. 
%
The first panel shows the dependence of the critical wave vectors on $\xi_9$, which is the coefficient of $(n_3\cdot v)^2$ in equation (\ref{H-def}). Using our coordinate system this dot product is independent of the azimuthal angle $\phi$, 
and therefore we have  that the terms in the critical wave vectors that are proportional to $\xi_9$ are also independent of $\phi$. 
The second panel shows the dependence of the critical wave vectors on $\xi_2$ which is the coefficient of $(n_3\cdot v)^2$. Both  the $\xi_9$ and $\xi_2$ terms have been included in previous calculations. In the third panel we show the dependence of the critical wave vectors on $\xi_6$, which is the coefficient of $(n_1\cdot v)(n_3\cdot v)$, and in the last panel we show the dependence on 
$\xi_8$, the coefficient of $(n_1\cdot v)^3(n_3\cdot v)$. 

\end{document}